\documentclass[nohyperref]{article}

\usepackage{microtype}
\usepackage{graphicx}
\usepackage{enumerate}
\usepackage[noend]{algpseudocode}
\usepackage{subfigure}
\usepackage{booktabs} 

\usepackage{hyperref}

\usepackage[accepted]{icml2023}

\usepackage{amsmath}
\usepackage{amssymb}
\usepackage{mathtools}
\usepackage{amsthm}
\usepackage{dsfont}
\usepackage[mathscr]{eucal}

\usepackage[capitalize,noabbrev]{cleveref}

\theoremstyle{plain}
\newtheorem{theorem}{Theorem}[section]
\newtheorem{proposition}[theorem]{Proposition}
\newtheorem{lemma}[theorem]{Lemma}

\theoremstyle{definition}

\theoremstyle{remark}

\usepackage[textsize=tiny]{todonotes}

\icmltitlerunning{On Distribution Dependent Sub-Logarithmic Query Time of Learned Indexing}

\DeclareMathAlphabet{\mathcal}{OMS}{cmsy}{m}{n}
\SetMathAlphabet{\mathcal}{bold}{OMS}{cmsy}{b}{n}

\begin{document}

\newcommand\normx[1]{\Vert#1\Vert}
\newcommand{\revision}[1]{#1}
\DeclareRobustCommand{\rchi}{{\mathpalette\irchi\relax}}
\newcommand{\irchi}[2]{\raisebox{\depth}{$#1\chi$}} 

\setlength{\abovedisplayskip}{0pt}
\setlength{\belowdisplayskip}{0pt}
\setlength{\abovedisplayshortskip}{0pt}
\setlength{\belowdisplayshortskip}{0pt}

\setlength{\textfloatsep}{4pt}
\setlength{\floatsep}{0pt}
\setlength{\intextsep}{0pt}
\setlength{\dbltextfloatsep}{0pt}
\setlength{\dblfloatsep}{0pt}
\setlength{\abovecaptionskip}{0pt}
\setlength{\belowcaptionskip}{0pt}

\twocolumn[
\icmltitle{On Distribution Dependent Sub-Logarithmic Query Time of Learned Indexing}




\begin{icmlauthorlist}
\icmlauthor{Sepanta Zeighami}{yyy}
\icmlauthor{Cyrus Shahabi}{yyy}
\end{icmlauthorlist}

\icmlaffiliation{yyy}{Univerisity of Southern California}

\icmlcorrespondingauthor{Sepanta Zeighami}{zeighami@usc.edu}
\icmlcorrespondingauthor{Cyrus Shahabi}{shahabi@usc}

\icmlkeywords{Machine Learning, ICML}

\vskip 0.3in
]



\printAffiliationsAndNotice{}  

\begin{abstract}
 A fundamental problem in data management is to find the elements in an array that match a query. Recently, learned indexes are being extensively used to solve this problem, where they learn a model to predict the location of the items in the array. They are empirically shown to outperform non-learned methods (e.g., B-trees or binary search that answer queries in $O(\log n)$ time) by orders of magnitude. However, success of learned indexes has not been theoretically justified. Only existing attempt shows the same query time of $O(\log n)$, but with a constant factor improvement in space complexity over non-learned methods, under some assumptions on data distribution. In this paper, we significantly strengthen this result, showing that under mild assumptions on data distribution, and the same space complexity as non-learned methods, learned indexes can answer queries in $O(\log\log n)$ expected query time. We also show that allowing for slightly larger but still near-linear space overhead, a learned index can achieve $O(1)$ expected query time. Our results theoretically prove learned indexes are orders of magnitude faster than non-learned methods, theoretically grounding their empirical success. 
\end{abstract}



\section{Introduction}
\vspace{-0.2cm}
It has been experimentally observed, but with little theoretical backing, that the problem of finding an element in an array has very efficient learned solutions \cite{galakatos2019fiting, kraska2018case, ferragina2020pgm, ding2020alex}. In this fundamental problem in data management, the goal is to find, given a query, the elements in the dataset that match the query (e.g., find the student with grade=$q$, for a number $q$, where ``grade=$q$'' is the query on a dataset of students). Assuming the query is on a single attribute (e.g., we filter students only based on grade), and that data is sorted based on this attribute, binary search finds the answer in $O(\log n)$ for an ordered dataset with $n$ records. Experimental results, however, show that learning a model (called a \textit{learned index} \cite{kraska2018case}) that predicts the location of the query in the array can provide accurate estimates of the query answers orders of magnitude faster than binary search (and other non-learned approaches). The goal of this paper is to present a theoretical grounding for such empirical observations.

\vspace{-0.1cm}More specifically, we are interested in answering exact match and range queries over a sorted array $A$. Exact match queries ask for the elements in $A$ exactly equal to the query $q$ (e.g., grade=$q$), while range queries ask for elements that match a range $[q, q']$ (e.g., grade is between $q$ and $q'$). Both queries can be answered by finding the index of the largest element in $A$ that is smaller than or equal to $q$, which we call the rank of $q$, \textit{rank}$(q)$. Range queries require the extra step, after obtaining rank$(q)$, of scanning the array sequentially from $q$ up to $q'$ to obtain all results. The efficiency of methods answering range and exact match  queries depends on the efficiency of answering rank$(q)$, which is the operation analyzed in the rest of this paper.

\begin{figure}
    \centering
    \includegraphics[width=0.8\columnwidth]{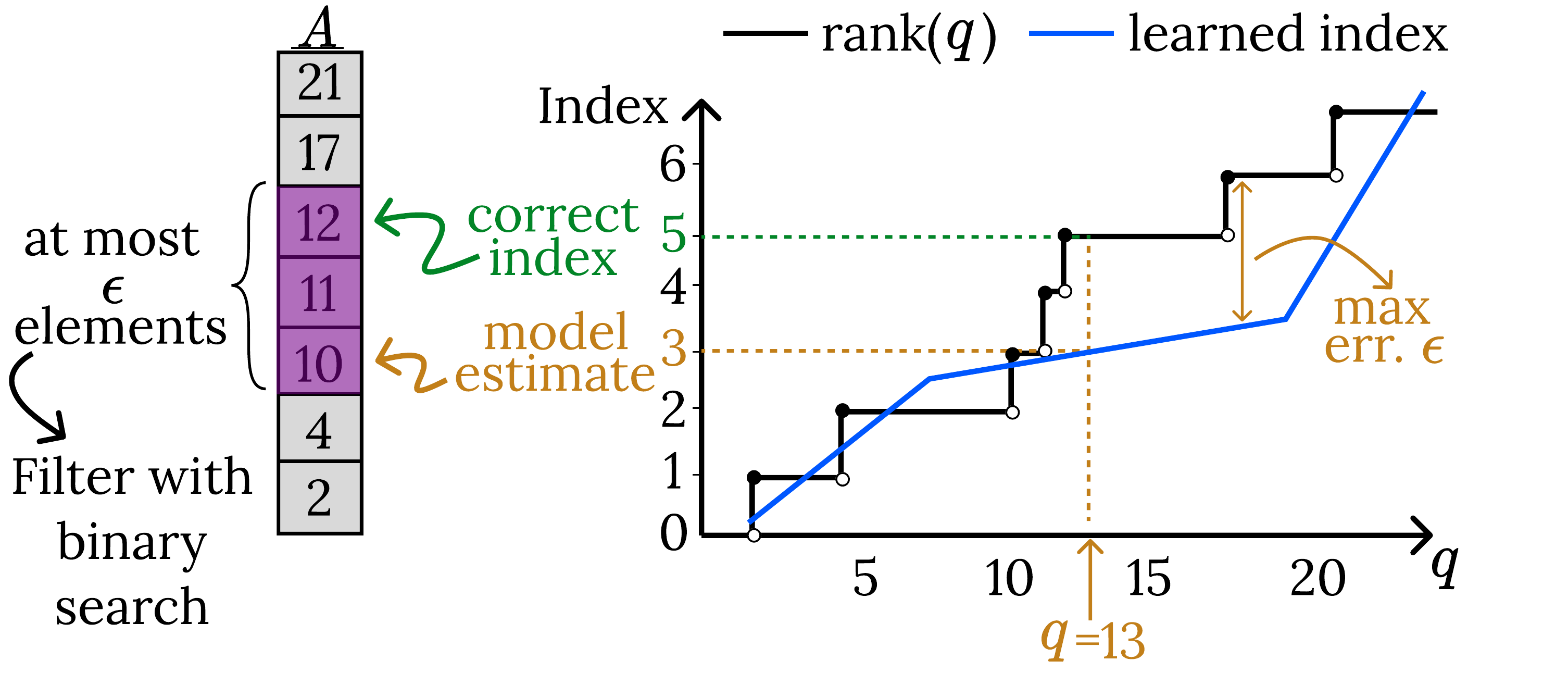}
    \vspace{-0.3cm}
    \caption{A learned index used to solve the rank problem.}
    \label{fig:learned_index_example}
\end{figure}

\vspace{-0.1cm}In the worst-case, and without further assumption on the data, binary search finds rank$(q)$ optimally, and in $O(\log n)$ operations. Materializing the binary search tree and variations of it, e.g., B-Tree \cite{bayer1970organization} and CSS-trees \cite{rao1998cache}, utilize caching and hardware properties to improve the performance in practice but theoretical number of operations remains $O(\log n)$ (we consider data in memory and not external storage).
On the other hand, \textit{learned indexes} have been empirically shown to outperform non-learned methods by orders of magnitude. Such approaches learn a model that predicts rank$(q)$. At query time, a model inference provides an estimate of rank$(q)$, and a local search is performed around the estimate to find the exact index. An example is shown in Fig.~\ref{fig:learned_index_example}, where for the query 13, the model returns index 3, while the correct index is 5. Then, assuming the maximum model error is $\epsilon$, a binary search on $\epsilon$ elements of $A$ within the model prediction (i.e., the purple sub-array in Fig.~\ref{fig:learned_index_example}) finds the correct answer. The success of learned models is attributed to exploiting patterns in the observed data to learn a small model that accurately estimates the rank of a query in the array. 

However, to date, no theoretical result has justified their superior practical performance. \citet{ferragina2020pgm} shows a worst-case bound of $O(\log n)$ on query time, the same as traditional methods, but experimentally shows orders of magnitude difference. The only existing result that shows any theoretical benefit to learned indexing is \citet{ferragina2020learned}, that shows constant factor better space utilization while achieving $O(\log n)$ query time under some assumptions on data distribution. The question remains whether theoretical differences, beyond merely constant factors, exist between learned and traditional approaches.

We answer this question affirmatively. We show that 
\vspace{-0.4cm}
\begin{enumerate}[(i)]
    \item Using the same space overhead as traditional indexes (e.g., a B-tree), and under mild assumptions on the data distribution, a learned index can answer queries in $O(\log\log n)$ operations on expectation, a significant and asymptotic improvement over the $O(\log n)$ of traditional indexes; 
   \vspace{-0.1cm}
 \item With the slightly higher but still near-linear space consumption $O(n^{1+\epsilon})$, for any $\epsilon>0$, a learned index can achieve $O(1)$ expected query time; and
   \vspace{-0.2cm}
    \item Under stronger assumptions on data distribution, we show that $O(\log\log n)$ expected query time is also possible with $O(1)$ space overhead ($O(1)$ space overhead is similar to performing binary search without building any auxiliary data structure). 
\end{enumerate}
\vspace{-0.3cm}
We present experiments showing these asymptotic bounds are achieved in practice. 

\vspace{-0.1cm}
These results show order of magnitude benefit in terms of expected query time, where the expectation is over the sampling of the data, and not worst-case query time
(which, unsurprisingly, is $O(\log n)$ in all cases). Intuitively, this means that although there may exist data instances where a learned index is as slow as binary search, for many data instances (and on expectation), it is fast and sub-logarithmic. Analyzing expected query time allows us to incorporate properties of the data distribution. Our results hold assuming certain distribution properties: query time in (i) and (ii) is achieved assuming bounded p.d.f of data distribution ((i) also assumes non-zero p.d.f), while (iii) assumes the c.d.f of data distribution is efficiently computable. 
Overall, data distribution had been previously hypothesized to be an important factor on the performance of a learned index (e.g., \citet{kraska2018case}). This paper shows how such properties can be used to analyze the performance of a learned index.


\section{Preliminaries and Related Work}\label{sec:prelim}
\vspace{-0.2cm}\subsection{Problem Definition}\vspace{-0.2cm}
\textbf{Setup}. We are given an array $A\subseteq \mathcal{D}^n$, consisting of $n$ elements, where $\mathcal{D}\subseteq\mathbb{R}$ is the domain of the elements. Unless otherwise stated, assume $\mathcal{D}=[0, 1]$; we discuss extensions to other bounded or unbounded domains in Sec.~\ref{sec:data_dist}. $A$ is sorted in ascending order, where $a_i$ refers to the $i$-th element in this sorted order and $A[i:j]$ denotes the sorted subarray containing \{$a_i$, ..., $a_{j}$\}. We assume $A$ is created by sampling $n$ i.i.d random variables and then sorting them, where the random variables follow a continuous distribution $\chi$, with p.d.f $f_\chi$ and c.d.f $F_\chi$. We use the notation $A\sim \chi$ to describe the above sampling procedure.

\textbf{Rank Problem}. Our goal is to answer the \textit{rank problem}: given the array $A$ and a query $q$, return the index $i^*= \sum_{i=1}^n I_{A[i]\leq q}$, 
where $I$ is the indicator function. $i^*$ is the index of the largest element no greater than $q$ and is 0 if no such element exists. Furthermore, if $q\in A$, $q$ will be at index $i^*+1$. We define the \textit{rank function} of an array $A$ as $r_A(q)=\sum_{i=1}^n I_{A[i]\leq q}$. The rank function takes a query as an input and outputs the answer to the rank problem. We drop the dependence on $A$ if it is clear from context and simply use $r(q)$. 

The rank problem is therefore the problem of designing a computationally efficient method to evaluate the function $r(q)$. Let $\hat{R}_A(q;\theta)$ be a function approximator, with parameters $\theta$ that correctly evaluates $r(q)$. The parameters $\theta$ of $\hat{R}_A$ are found at a preprocessing step and are used to perform inference at query time. Let $T(\hat{R}_A, q)$ be the number of operations performed by $\hat{R}_A$ to answer the query $q$ and let $S(\hat{R}_A)$ be the space overhead of $\hat{R}_A$, i.e., the number of bits required to store $\theta$ (note that $S(\hat{R}_A)$ does not include storage required for the data itself, but only considers the overhead of indexing). We study the \textit{expected query time} of any query $q$ as $\mathds{E}_{A\sim \chi}[T(\hat{R}_A, q)]$, and the \textit{expected space overhead} as $\mathds{E}_{A\sim \chi}[S(\hat{R}_A)]$. In our analysis of space overhead, we assume integers are stored using their binary representation so that $k$ integers that are at most $M$ are stored in $O(k\log M)$ bits (i.e., assuming no compression).

\textbf{Learned indexing}. A \textit{learned indexing} approach solves the rank problem as follows. A function approximator (e.g., neural network or a piecewise linear approximator) $\hat{r}_A(q; \theta)$ is first learned that approximates $r$ up to an error $\epsilon$, i.e., $\normx{\hat{r}_A-r}_\infty \leq \epsilon$. Then, at the step called \textit{error correction}, another function, $h(i, q)$, takes the estimate $i=\hat{r}_A(q;\theta)$ and corrects the error, typically by performing a binary search (or exponential search when $\epsilon$ is not known a priori \cite{ding2020alex}) on the array, $A$. That is, given that the estimate $\hat{r}_A$ is within  $\epsilon$ of the true index of $q$ in $A$, a binary search on the $2\epsilon$ element in $A$ that are within $\epsilon$ of $\hat{r}_A(q; \theta)$ finds the correct answer. Letting $\hat{R}_A(q;\theta)=h(\hat{r}_A(q, \theta), q)$, we obtain that for any function approximator, $\hat{r}_A$ with non-zero error $\epsilon$, we can obtain an exact function with expected query time of $\mathds{E}_{A\sim \chi}[T(\hat{r}_A, q)]+O(\log \epsilon)$ and space overhead of $\mathds{E}_{A\sim \chi}[S(\hat{r}_A)]$ since binary search requires no additional storage space. In this paper, we show the existence of function approximators, $\hat{R}_A$ that can achieve sub-logarithmic query time with various space overheads.  


\vspace{-0.2cm}\subsection{Related Work}\vspace{-0.2cm}
\textbf{Learned indexes}. The only existing work theoretically studying a learned index is \citet{ferragina2020learned}. It shows, under assumptions on the gaps between the keys in the array, as $n\rightarrow\infty$ and almost surely, one can achieve logarithmic query time with a learned index with a constant factor improvement in space consumption over non-learned indexes. We significantly strengthen this result, showing \textit{sub-logarithmic} expected query time under various space overheads. Our assumptions are on the data distribution itself which is more natural than assumption on the gaps, and our results hold for any $n$ (and not as $n\rightarrow\infty$). Though scant in theory, learned indexes have been extensively utilized in practice, and various modeling choices have been proposed under different settings, e.g., \citet{galakatos2019fiting, kraska2018case, ferragina2020pgm, ding2020alex} to name a few. Our results use a hierarchical model architecture, similar to Recursive Model Index (RMI) \cite{kraska2018case} and piecewise approximation similar to Piecewise Geometric Model index (PGM) \cite{ferragina2020pgm} to construct function approximators with sub-logarithmic query time. 

\textbf{Non-Learned Methods}. Binary search trees,  B-Trees \cite{bayer1970organization} and many other variants \cite{rao1998cache, lehman1985study, bayer1972symmetric}, exist that solve the problem in $O(\log n)$ query time, which is the best possible in the worst case in comparison based model \cite{navarro2020predecessor}. The space overhead for such indexes is $O(n\log n)$ bits, as they have $O(n)$ nodes and each node can be stored in $O(\log n)$ bits. We also note in passing that if we limit the domain of elements to a finite integer universe and do not consider range queries, various other time/space trade-offs are possible \cite{puatracscu2006time}, e.g., using hashing \cite{fredman1984storing}.


\if 0\begin{table}[]
    \centering
    \begin{tabular}{c c c}
    \toprule
        \textbf{Space} & \textbf{Time (Learned)}  & \textbf{Time (Non-Learned)} \\
    \midrule
        $n^{1+\epsilon}$ & $1$ [ours] & $\frac{\log l}{\log\log l}$ [STOC’99]\\
        $n\log^{O(1)}n$ & $\log\log n$ [ours] & $\log l$ [FOCS’75]\\
        $1$ & $\log\log n$ [ours] & $\log n$ [binary search] \\
    \bottomrule
    \end{tabular}
    \caption{Best known bounds on expected query time given a space overhead (all expressions are asymptotic, e.g., 1 means $O(1)$). $l$ can be $O(n)$}
    \label{tab:results}
\end{table}
\fi
\vspace{-0.4cm}
\section{Asymptotic Behaviour of Learned Indexing}\label{sec:results_overview}\vspace{-0.2cm}

\subsection{Constant Time and Near-Linear Space} \vspace{-0.1cm}
We first consider the case of constant query time. 
\vspace{-0.1cm}\begin{theorem}\label{theorem:superlinear}
    Suppose the p.d.f, $f_\chi(x)$, is bounded, i.e., $f_\chi(x) \leq \rho$ for all $x\in \mathcal{D}$, where $\rho<\infty$. There exists a learned index with space overhead \revision{$O(\rho^{1+\epsilon}n^{1+\epsilon})$}, for any $\epsilon>0$, with expected query time of \revision{$O(\log \frac{1}{\epsilon})$} operations for any query. \revision{$\rho$ is a constant independent of $n$, and for any constant $\epsilon$, asymptotically in $n$, space overhead is $O(n^{1+\epsilon})$ and expected query time is $O(1)$.}
\end{theorem}
\vspace{-0.2cm}
The theorem shows the surprising result that we can in fact achieve constant query time with a learned index of size $O(n^{1+\epsilon})$. Although the space overhead is near-linear, this overhead is asymptotically larger than the overhead of traditional indexes (with overhead $O(n\log n)$) and thus the query time complexities are not directly comparable.  

\begin{figure}
    \centering
    \begin{minipage}{0.49\columnwidth}
        \centering
        \includegraphics[width=\textwidth]{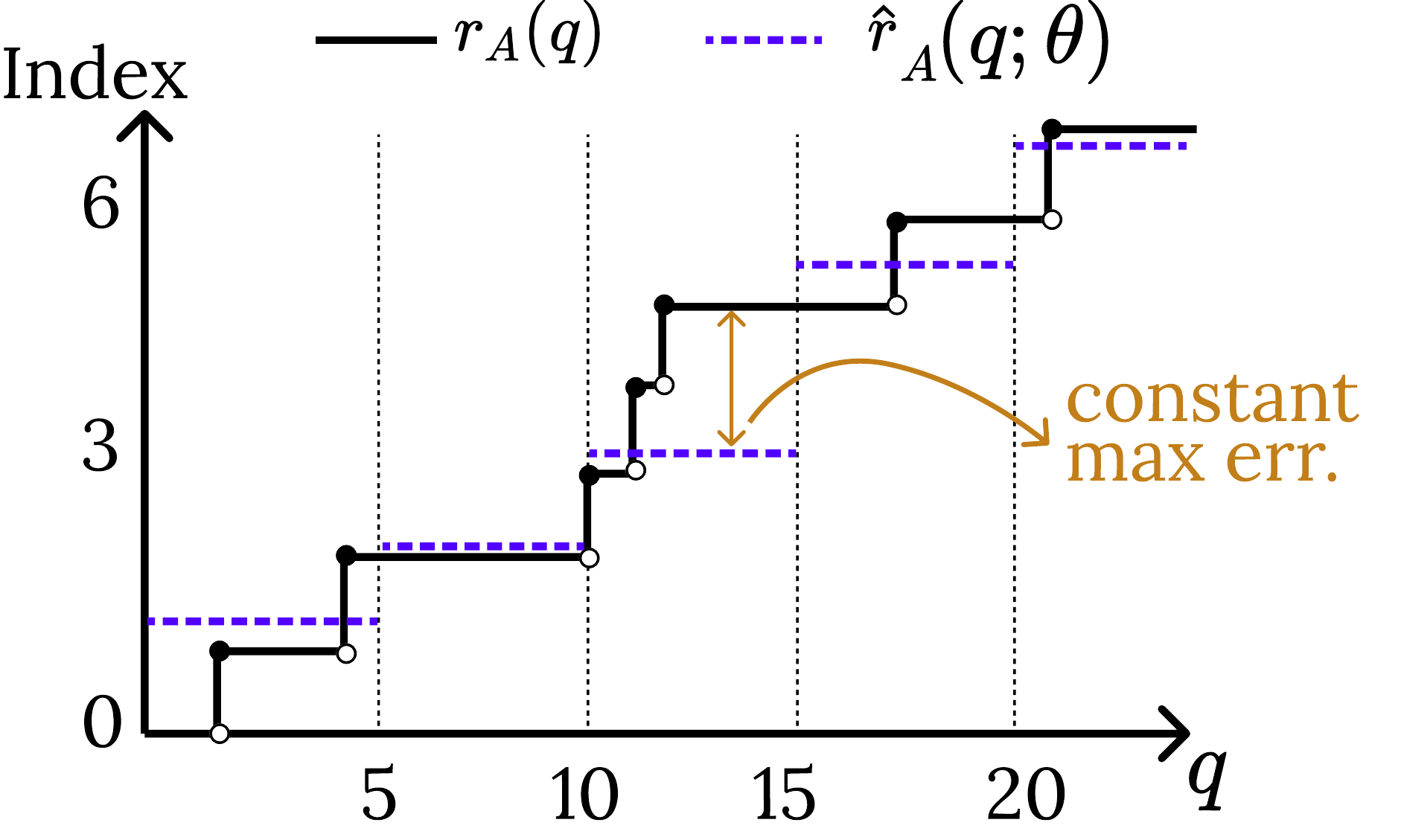}
    \vspace{-0.7cm}
        \caption{Approximation with a piecewise constant function }
        \label{fig:unif_appx}
    \end{minipage}
    \hfill
    \begin{minipage}{0.49\columnwidth}
        \centering
        \includegraphics[width=\textwidth]{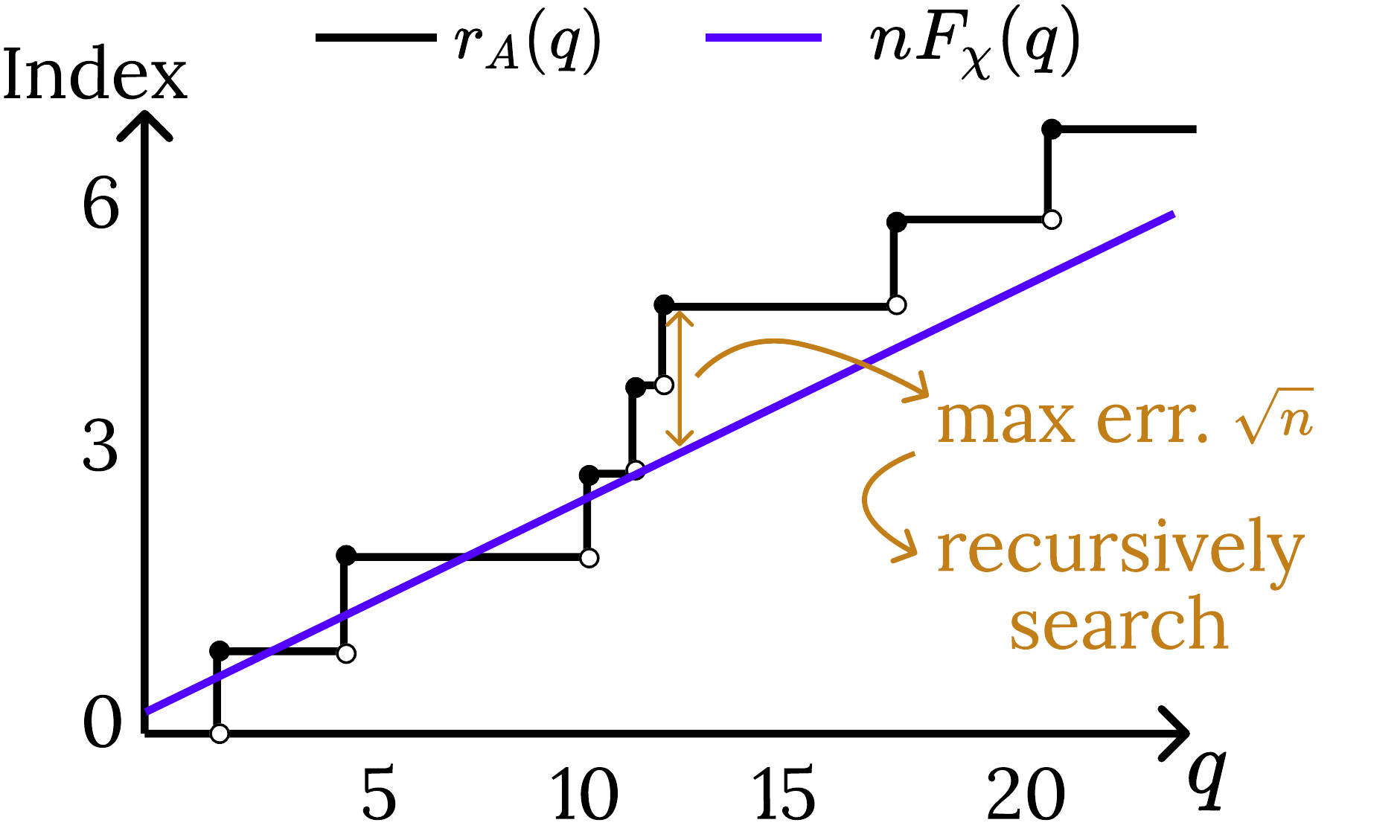}
        \vspace{-0.7cm}
        \caption{Approximation with c.d.f}
        \label{fig:cdf_appx}
    \end{minipage}
\end{figure}

Interestingly, the function approximator that achieves the bound in Theorem~\ref{theorem:superlinear} is a simple piecewise constant function approximator, which can be seen as a special case of the PGM model that uses piece-wise linear approximation \cite{ferragina2020pgm}. Our function approximator is constructed by uniformly dividing the space into $k$ intervals and for each interval finding a constant that best approximates the rank function in that interval. Such a function approximator is shown as $\hat{r}_A(q;\theta)$ in Fig.~\ref{fig:unif_appx} for $k=5$. Obtaining constant query time requires such a function approximator to have constant error. It is, however, non-obvious why and when only $O(n^{1+\epsilon})$ pieces will be sufficient on expectation to achieve constant error. In fact, for the worst-case (and not the expected case), for a heavily skewed dataset, achieving constant error would require an arbitrarily large $k$, as noted by \citet{kraska2018case}. 

However, Theorem~\ref{theorem:superlinear} shows as long as the p.d.f. of the data distribution is bounded, $O(n^{1+\epsilon})$ pieces will be sufficient for constant query time on expectation. Intuitively, the bound on the p.d.f. is used to argue that the number of data points sampled in a small region is not too large, which is in turn used to bound the error of the function approximation. 

\revision{Finally, dependence on $\rho$ in Theorem~\ref{theorem:superlinear} is expected, as performance of learned indexes depends on the dataset characteristics. $\rho$ captures such data dependencies, showing that such data dependencies only affect space overhead by a constant factor. From a practical perspective, our experiments in Sec.~\ref{sec:exp:real} show that for many commonly used real-world benchmarks for learned indexes, trends predicted by Theorem~\ref{theorem:superlinear} hold with $\rho=1$. However, Sec.~\ref{sec:exp:real} also shows that for datasets where learned indexes are known to perform poorly, we observe large values of $\rho$. Thus, $\rho$ can be used to explain why and when learned indexes perform well or poorly in practice.}


\vspace{-0.2cm}
\subsection{Log-Logarithmic Time and Constant Space}\vspace{-0.2cm}
Requiring constant query time, as in the previous theorem, can be too restrictive. 
Allowing for slightly larger query time, we have the following result. 

 \begin{theorem}\label{theorem:constant}
    Suppose c.d.f of data distribution $F_\chi(x)$ can be evaluated exactly with $O(1)$ operations and $O(1)$ space overhead. There exists a learned index with space overhead $O(1)$, where for any query $q$, the expected query time is $O(\log \log n)$ operations.
\end{theorem}

The result shows that we can obtain $O(\log\log n)$ query time if the c.d.f of the data distribution is easy to compute. This is the case for the uniform distribution (whose c.d.f is a straight line), or more generally any distribution with piece-wise polynomial c.d.f. In this regime, we only utilize constant space, and thus our bound is comparable with performing a binary search on the array, which takes $O(\log n)$ operations, showing that the learned approach enjoys an order of magnitude theoretical benefit.

Our model of the rank function is $n\times F_\chi$, where $F_\chi$ is the c.d.f of the data distribution. As Fig.~\ref{fig:cdf_appx} shows, our search algorithm proceeds recursively, at each iteration reducing the search space by around $\sqrt{n}$. Intuitively, the $\sqrt{n}$ is due to the Dvoretzky-Kiefer-Wolfowitz (DKW) bound \cite{massart1990tight}, which is used to show that with high probability the answer to a query, $q$ is within $\sqrt{n}$ of $nF_\chi(q)$. Reducing the search space, $s$, by roughly $\sqrt{s}$ at every level by recursively applying DKW, we obtain the total search time of $O(\log\log n)$ (note that binary search only reduces the search space by a factor of $2$ at every iteration). 


\begin{figure}
    \centering
    \includegraphics[width=\columnwidth]{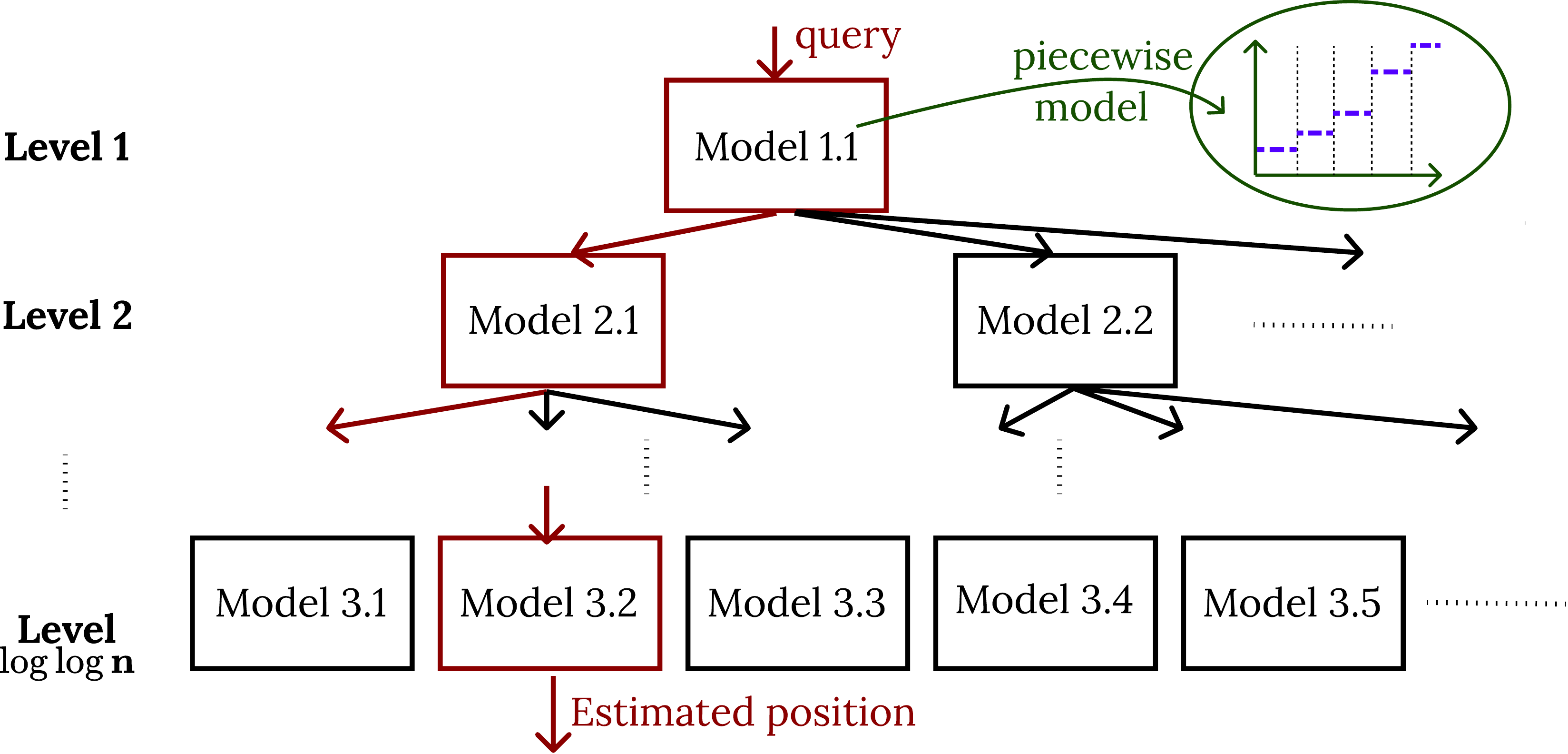}
    \vspace{-0.7cm}
    \caption{RMI of height $\log\log n$ with piecewise constant models}
    \label{fig:rmi}
\end{figure}

\vspace{-0.2cm}\subsection{Log-Logarithmic Time and Quasi-Linear Space}\vspace{-0.2cm} 
Finally, we show that the requirement of Theorem~\ref{theorem:constant} on the c.d.f. is not necessary to achieve $O(\log\log n)$ query time, provided quasi-linear space overhead is allowed. The following theorem shows that a learned index can achieve $O(\log\log n)$ query time under mild assumptions on the data distribution and utilizing quasi-linear space.

\begin{theorem}\label{theorem:linear}
    Suppose p.d.f of data distribution $f_\chi(x)$ is bounded and more than zero, i.e., $\rho_1 \leq f_\chi(x) \leq \rho_2$ for all $x\in \mathcal{D}$, where $\rho_1>0$ and $\rho_2<\infty$. There exists a learned index with expected query time equal to $O(\log \log n)$ operations and space overhead \revision{$O(\frac{\rho_2}{\rho_1}n\log n)$}, for any query. \revision{Specifically, $\frac{\rho_2}{\rho_1}$ is a constant independent of $n$, so that, asymptotically in $n$, space overhead is $O(n\log n)$}.
\end{theorem}

This regime takes space similar to data size, and is where most traditional indexing approaches lie, e.g., binary trees and B-trees, where they need $O(n\log n)$ storage (the $\log n$ is due to the number of bits needed to store each node content) and achieve $O(\log n)$ query time. 

The learned index that achieves the bound in Theorem~\ref{theorem:linear} is an instance of the Recursive Model Index (RMI) \cite{kraska2018case}. Such a learned index defines a hierarchy of models, as shown in Fig.~\ref{fig:rmi}. Each model is used to pick a model in the next level of the tree until a model in the leaf level is reached, whose prediction is the estimated position of the query in the array. Unlike RMI in \cite{kraska2018case}, its height or size of the model within each node is not constant and set based on data size. 

Intuitively, the hierarchy of models is a materialization of a search tree based on the recursive search used to prove Theorem~\ref{theorem:constant}. At any level of the tree, if the search space is $s$ elements (originally, $s=n$) a model is used to reduce the search space to roughly $\sqrt{s}$. It is however non-trivial why and when such a model should exist across all levels and how large the model should be. We use the relationship between the rank function and the c.d.f (through DKW bound), and the properties of the data distribution to show that a model of size around $\sqrt{s}$ is sufficient with high probability. Note that models at lower levels of the hierarchy approximate the rank function only over subsets of the array, but with increasingly higher accuracy. A challenge is to show that such an approximability result holds across all models and all subsets of the array, which is why a lower bound on the p.d.f. is needed in this theorem. 

\revision{Similar to $\rho$ in Theorem~\ref{theorem:superlinear}, $\rho_1$ and $\rho_2$ capture data characteristics in Theorem~\ref{theorem:linear}, showing constant factor dependencies on the model size. Our experiments in Sec.~\ref{sec:exp:real} show that for most commonly used real-world benchmarks for learned indexes, trends predicted by Theorem~\ref{theorem:linear} hold with $\frac{\rho_2}{\rho_1}=1$. However, Sec.~\ref{sec:exp:real} also shows that for datasets where learned indexes are known to perform poorly, $\frac{\rho_2}{\rho_1}$ is large, so that $\frac{\rho_2}{\rho_1}$ can be used to explain why and when learned indexes perform well or poorly in practice.}

\vspace{-0.2cm}\subsection{Distributions with Other Domains}\label{sec:data_dist}\vspace{-0.2cm}
So far, our results assume that the domain of data distribution is [0, 1]. \revision{The result can be extended to distributions with other bounded domains, $[r, s]$ for $r,s\in \mathbb{R}$, $r<s$, by standardizing $\chi$ as $\frac{\chi-r}{s-r}$. This transformation scales p.d.f of $\chi$ by $s-r$. Note that scaling the p.d.f does not affect Theorem~\ref{theorem:linear}, since both $\rho_1$ and $\rho_2$ will be scaled by $s-r$, yielding the same ratio $\frac{\rho_2}{\rho_1}$. On the other hand, $\rho$ in Theorem~\ref{theorem:superlinear} will be scaled by $s-r$. } Overall, bounded domain can be true in many scenarios, as the data can be from some phenomenon that is bounded, e.g., age, grade, data over a period of time. Next, we extend our results to distributions with unbounded domains.

\begin{lemma}\label{lemma:subexp}
    Suppose a learned index, $\hat{R}$, achieves expected query time $t(n)$ and space overhead $s(n)$ on distributions with domain $[0, 1]$ and bounded (and non-zero) p.d.f. There exists a learned index, $\hat{R}'$ with expected query time $t(n)+1$ and space overhead $O(s(n)\log n)$ on any sub-exponential distribution with bounded (and non-zero) p.d.f. 
\end{lemma}
\vspace{-0.2cm}
Combining Lemma~\ref{lemma:subexp} with Theorems \ref{theorem:superlinear} and \ref{theorem:linear}, our results cover various well-known distributions, e.g., Gaussian, squared of Gaussian and exponential distributions. 

Proof of lemma~\ref{lemma:subexp} builds the known learned index for bounded domains on $\log n$ different bounded intervals. 
This achieves the desired outcome due to the tail behavior of sub-exponential distributions (i.e., for distributions with tail at most as heavy as exponential, see \citet{vershynin2018high} for definition). The tail behaviour allows us to, roughly speaking, assume that the domain of the function is $O(\log n)$, because observing points outside this range is unlikely. 
We note that other distributions with unbounded domain can also be similarly analyzed based on their tail behaviour, with heavier tails leading to higher space consumption. 

\vspace{-0.4cm}\section{Proofs}\label{sec:proofs}\vspace{-0.2cm}
Proofs of the theorems are all constructive. PCA Index (Sec.~\ref{sec:pcf}) proves Theorem~\ref{theorem:superlinear}, RDS algorithm  proves Theorem~\ref{theorem:constant} and RDA Index proves Theorem~\ref{theorem:linear}. 
Without loss of generality, we assume the bounded domain $\mathcal{D}$ is $[0, 1]$. The proof for the unbounded domain case (i.e., Lemma~\ref{lemma:subexp}) is deferred to Appendix~\ref{appx:proofs}. Proof of technical lemmas stated throughout this section can also be found in Appendix~\ref{appx:proofs}.

\vspace{-0.4cm}
\subsection{Proof of Theorem~\ref{theorem:superlinear}: PCA Index}\label{sec:pcf}\vspace{-0.2cm}
We present and analyze \textit{Piece-wise Constant Approximator (PCA) Index} that proves Theorem~\ref{theorem:superlinear}. 
\vspace{-0.2cm}\subsubsection{Approximating Rank Function}\label{sec:PCA:app}
We show how to approximate the rank function $r$ with a function approximator $\hat{r}$. To achieve constant query time, approximation error should be a constant independent of $n$ with high probability, and we also should be able to evaluate $\hat{r}$  in constant time. 

Lemma~\ref{lemma:const_error} shows these properties hold for a piece-wise constant approximation to $r$. Such a function is presented in Alg.~\ref{alg:pcf} (and an example was shown in Fig.~\ref{fig:unif_appx}). Alg.~\ref{alg:pcf} uniformly divides the function domain into $k$ intervals, so that the $i$-th constant piece is responsible for the interval $I_i=[i\times \frac{1}{k}, (i+1)\times \frac{1}{k}]$. Since $r(q)$ is a non-decreasing function, the constant with the lowest  infinity norm error approximating $r$ over $I_i$ is $\frac{1}{2}(r(\frac{i}{k})+r(\frac{i+1}{k}))$ (line~\ref{alg:pcf:best_piece}).  Let $\hat{r}_k$ be the function returned by \textsc{PCF($A$, $k$, $0$, $1$)}.
\begin{lemma}\label{lemma:const_error}
    Under the conditions of Theorem~\ref{theorem:superlinear} and for $k\geq n^{1+\epsilon}\rho^{1+\frac{\epsilon}{2}}$, the error of $\hat{r}_k$ is bounded as 
$$
\mathds{P}(\normx{\hat{r}_k-r}_\infty\geq \frac{2}{\epsilon}+1)\leq \frac{1}{n}.
$$
\end{lemma}
\textit{Proof of Lemma~\ref{lemma:const_error}}. Let $e_i=\sup_{x\in I_i}|\hat{r}(x;\theta)-r(x)|$ be the maximum error in the $i$-th piece of $\hat{r}$. $e_i$ can be bounded by the number of points sampled in $I_i$ as follows. 
\begin{proposition}\label{prop:bound_err_by_samples}
    Let $s_i=|\{j|a_j\in I_i\}|$ be the number of points in $A$ that are in $I_i$. We have $e_i\leq s_i$ 
\end{proposition}
Using Prop.~\ref{prop:bound_err_by_samples}, we have $\normx{\hat{r}-r}_\infty \leq \max_{i\in \{1, ..., k\}} s_i$. Prop.~\ref{prop:bound_err_by_samples} is a simple fact that relates approximation error to statistical properties of data distribution. Define $s_{max}=\max_{i\in \{1, ..., k\}} s_i$ and observe that $s_{max}$ is a random variable denoting the maximum number of points sampled per interval, across $k$ equi-length intervals. The following lemma shows that we can bound $s_{max}$ with a constant and with probability $\frac{1}{n}$, as long as $k$ is near-linear in $n$. 

\begin{lemma}\label{lemma:prob_poins_in_piece}
    For any $c$ with $c\geq 3$, and if $k\geq n^{1+\frac{2}{c-1}}\rho^{1+\frac{1}{c-1}}$ we have $\mathds{P}(s_{max}\geq c)\leq \frac{1}{n}$.
\end{lemma}

Setting $c=\frac{2}{\epsilon}+1$, we see $k\geq n^{1+\epsilon}\rho^{1+\frac{\epsilon}{2}}$ holds, so that Lemma~\ref{lemma:prob_poins_in_piece} together with Prop.~\ref{prop:bound_err_by_samples} prove Lemma~\ref{lemma:const_error}.  \qed




\begin{algorithm}[t]
\begin{algorithmic}[1]
\Require A sorted array $A$, number of pieces $k$, approximation domain lower and upper bounds $l$ and $u$
\Ensure Piecewise constant approximation of $r$ over $[l, u]$
\Procedure{PCF}{$A$, $k$, $l$, $u$}
    \State $P\leftarrow$ array of length $k$ storing the pieces
    \State $\alpha\leftarrow \frac{(u-l)}{k}$
    \State $\delta\leftarrow 0$
    \For{$i\leftarrow0$ to $k$}
        \State $P[i]\leftarrow \frac{1}{2}(r_A(l+ \alpha i)+r_A(l+ \alpha (i+1))$
        \label{alg:pcf:best_piece}
        \State $\delta_{curr}\leftarrow \bigl\lceil\frac{1}{2}(r_A(l+ \alpha (i+1))-r_A(l+ \alpha i))\bigr\rceil$
        \State $\delta\leftarrow \max\{\delta, \delta_{curr}\}$
    \EndFor
    \Return $P$, $\delta$
\EndProcedure
\caption{PCA Index Construction}\label{alg:pcf}
\end{algorithmic}
\end{algorithm}

\subsubsection{Index Construction and Querying}\label{sec:unif_pcf:const}
Let $k= \lceil n^{1+\frac{\epsilon}{2}}\rho^{1+\frac{\epsilon}{4}}\rceil$. We use \textsc{PCF($A$, $k$, $0$, $1$)} to obtain $\hat{r}_k$ and $\delta$, where $\delta$ is the maximum observed approximation error. As Alg.~\ref{alg:pcf} shows, $\hat{r}_k$ can be stored as an array, $P$, with $k$ elements. 
To perform a query, the interval, $i$, a query falls into is calculated as $i=\lfloor qk\rfloor$ and the constant responsible for that interval, $P[i]$, returns the estimate. Given maximum error $\delta$, we perform a binary search on the subarray $A[l:u]$, for $l=P[i]-\delta$ and $u=P[i]+\delta$ to obtain the answer.



\vspace{-0.2cm}
\subsubsection{Complexity Analysis}\label{sec:unif_pcf:analysis}
$P$ has $O(n^{1+\frac{\epsilon}{2}})$ entries, and each can be stored in $O(n^{\frac{\epsilon}{2}})$. Thus, total space complexity is $O(n^{1+\epsilon})$. 
%
%
Regarding query time, the number of operations needed to evaluate $\hat{r}_k$ is constant. Thus, the total query time of the learned index is $O(\log \delta)$. 
%
Lemma~\ref{lemma:const_error} bounds $\delta$, so that the query time for any query is at most $\log (\frac{4}{\epsilon}+1)$ with probability at least $1-\frac{1}{n}$ and at most $\log n$ with probability at most $\frac{1}{n}$. Consequently, the expected query time is at most $O(\log (\frac{4}{\epsilon}+1)\times(1-\frac{1}{n})+\log n\times \frac{1}{n})$ which is $O(1)$ for any constant $\epsilon>0$. \qed

\vspace{-0.2cm}
\subsection{Proof of Theorem~\ref{theorem:constant}: RDS Algorithm}\label{sec:constant_space}
\vspace{-0.2cm}
We present and analyze \textit{Recursive Distribution Search (RDS)} Algorithm that proves Theorem~\ref{theorem:constant}. 

\vspace{-0.3cm}
\subsubsection{Approximating Rank Function}\label{sec:rds:app}
\vspace{-0.2cm}
We approximate the rank function using the c.d.f of the data distribution, which conditions of Theorem~\ref{theorem:constant} imply is easy to compute. As noted by \citet{kraska2018case}, rank$(q)=nF_n(q)$, where $F_n$ is the empirical c.d.f. Using this together with DKW bound \cite{massart1990tight}, we can establish that rank($q$) is within error $\sqrt{n}$ of $nF_\chi$ with high probability. However, error of $\sqrt{n}$ is too large: error correction to find rank$(q)$ would require $O(\log \sqrt{n})=O(\log n)$ operations. 

Instead, we recursively improve our estimate by utilizing information that becomes available from observing elements in the array. After observing two elements, $a_i$ and $a_j$ in $A$ ($i<j$), we update our knowledge of the distribution of elements in $A[i+1:j-1]$ as follows. Define $F_{\chi}^{i, j}(x)=\frac{F_\chi(x)-F_\chi(a_i)}{F_\chi(a_j)-F_\chi(a_i)}$. 
Informally, any element $X$ in $A[i+1:j-1]$ is a random variable sampled from $\chi$ and knowing the value of $a_i$ and $a_j$ implies that $X\in [a_i, a_j]$, so that the conditional c.d.f of $X$ is $$\mathop{\mathds{P}}_{X\sim\chi}(X\leq x|a_i\leq X\leq a_j)=F_{\chi}^{i, j}(x).$$ We then use DKW bound to show $F_{\chi}^{i, j}$ is a good estimate of the rank function for the subarray $A[i+1:j-1]$, defining the rank function for the subarray $A[i+1:j-1]$ as $r^{i, j}(q)=\sum_{z=i+1}^{j-1} I_{a_z\leq q}$. \revision{Formally, the following lemma shows that given observations $A[i]=a_i$ and $A[j]=a_j$ the elements of $A[i+1:j-1]$ are i.i.d random variables with the conditional c.d.f $F_{\chi}^{i, j}(x)$ and uses the DKW bound to bound the approximation error of using the conditional c.d.f to approximate the conditional rank function.}

\begin{lemma}\label{lemma:dkw}
 Consider two indexes $i, j$, where $1\leq i < j\leq n$ and $a_i<a_j$. Let $k=j-i-1$. For $k\geq 2$, 
 we have
\begin{align*}
\mathds{P}(\sup_{x}|r^{i, j}(x)-kF^{i, j}_\chi(x)|\geq \sqrt{0.5k\log\log k})\leq \frac{1}{\log k}.
\end{align*}
\end{lemma}

\begin{algorithm}[t]
\begin{algorithmic}[1]
\Require A sorted array $A$ of size $n$ searched from index $i$ to $j$, a query $q$
\Ensure Rank of $q$ in $A[i:j]$
\Procedure{Search}{$A$, $q$, $i$, $j$}
\State $k\leftarrow j-i-1$
\If{$k<25$}\label{alg:cdf_search:base1}
\State \Return $i$-1+\textsc{BinarySearch}($A$, $q$, $i$, $j$)
\EndIf
\If{$a_i> q$} \Return $0$\label{alg:cdf_search:base2}
\EndIf
\If{$a_i= q$} \Return $1$
\EndIf
\If{$a_j\leq q$}
 \Return $j-i+1$\label{alg:cdf_search:base4}
\EndIf
\State $\hat{i} \leftarrow i+1 + k\times F_\chi^{i, j} (q)$\label{alg:search_cdf_index}
\State $r\leftarrow\sqrt{0.5k\log\log k}$
\State $l\leftarrow \lfloor \hat{i}-r\rfloor$
\State $u\leftarrow \lceil \hat{i}+r\rceil$\label{alg:search_cdf_range_end}
\If{$a_l> q$  \textbf{or}  $a_r< q$}\label{alg:search_cdf_check}
\State \Return $i-1+$\textsc{BinarySearch}($A$, $q$, $i$, $j$)
\EndIf
\State \Return $l-1+$ \textsc{Search}($A$, $q$,  $l$, $u$)
\EndProcedure
\caption{Recursive Distribution Search Algorithm}\label{alg:search_with_distribution}
\end{algorithmic}
\end{algorithm}

\vspace{-0.5cm}
\subsubsection{Querying}\label{sec:rds:query}\vspace{-0.2cm}
We use Lemma~\ref{lemma:dkw} to recursively search the array. At every iteration, the search is over a subarray $A[i:j]$ (initially, $i$=1 and $j=n$). We observe the values of $a_i$ and $a_j$ and use Lemma~\ref{lemma:dkw} to estimate which subarray is likely to contain the answer to the query. This process is shown in Alg.~\ref{alg:search_with_distribution}. In lines~\ref{alg:cdf_search:base2}-\ref{alg:cdf_search:base4} the algorithm observes $a_i$ and $a_j$ and attempts to answer the query based on those two observations. If it cannot, lines~\ref{alg:search_cdf_index}-\ref{alg:search_cdf_range_end} use Lemma~\ref{lemma:dkw} and the observed values of $a_i$ and $a_j$ to estimate which subarray contains the answer. Line~\ref{alg:search_cdf_check} then checks if the estimated subarray  is correct, i.e., if the query does fall inside the estimated subarray. If the estimate is correct, the algorithm recursively searches the subarray. Otherwise, the algorithm exits and performs binary search on the current subarray. Finally, line~\ref{alg:cdf_search:base1} exits when the size of the dataset is too small. The constant 25 is chosen for convenience of analysis (see Sec.~\ref{sec:const_space:analysis}). 


\vspace{-0.2cm}
\subsubsection{Complexity Analysis}\label{sec:const_space:analysis}
\vspace{-0.2cm}
To prove Theorem~\ref{theorem:constant}, it is sufficient to show that expected query time of Alg.~\ref{alg:search_with_distribution} is $O(\log\log n)$ for any query. The algorithm recursively proceeds. At each recursion level, the algorithm performs a constant number of operations unless it exits to perform a binary search. Let the depth of recursion be $h$ and let $k_i$ be the size of the subarray at the $i$-th level of recursion (so that binary search at $i$-th level takes $O(\log k_i)$). Let $B_i$ denote the event that the algorithm exits to perform binary search at the $i$-th iteration. Thus, for any query $q$, the expected number of operations is $$\mathds{E}_{A\sim \chi}[T(\hat{r}, q)]=\sum_{i=1}^{h} c_1+c_2\mathds{P}(B_i,\bar{B}_{i-1}, .... \bar{B}_1)\log k_i$$ for constants $c_1$ and $c_2$. Note that 
$\mathds{P}(B_i,\bar{B}_{i-1}, .... \bar{B}_1)\leq \mathds{P}(B_i|\bar{B}_{i-1}, .... \bar{B}_1)$, 
where $\mathds{P}(B_i|\bar{B}_{i-1}, .... \bar{B}_1)$ is the probability that the algorithm reaches $i$-th level of recursion and exits. By Lemma~\ref{lemma:dkw}, this probability bounded by $\frac{1}{\log k_i}$. Thus $\mathds{E}_{A\sim \chi}[T(\hat{r}, q)]$ is $O(h)$.

To analyze the depth of recursion, recall that at the last level, the size of the array is at most 25. Furthermore, at every iteration the size of the array is reduced to at most $2\sqrt{0.5 n \log\log n}+2$. For $n\geq 25$, $2\sqrt{0.5n\log\log n}+2\leq n^{\frac{3}{4}}$, so that the size of the array at the $i$-th recursions is at most $n^{(\frac{3}{4})^i}$ and the depth of recursion is $O(\log\log n)$. Thus, the expected total time is $O(\log\log n)$\qed.

\vspace{-0.2cm}
\subsection{Proof of Theorem~\ref{theorem:linear}: RDA Index}\label{sec:rdf}
\vspace{-0.2cm}
We present and analyze \textit{Recursive Distribution Approximator (RDA)} Index that proves Theorem~\ref{theorem:linear}. 

\vspace{-0.2cm}
\subsubsection{Approximating Rank Function}\label{sec:rda:app}
\vspace{-0.2cm}
We use ideas from Theorems~\ref{theorem:superlinear} and \ref{theorem:constant} to approximate the rank function.  We use Alg.~\ref{alg:search_with_distribution} as a blueprint, but instead of the c.d.f, we use a piecewise constant approximation to the rank function. If we can efficiently approximate the rank function  for subarray $A[i-1:j+1]$, $r^{i, j}$, to within accuracy $O(\sqrt{k\log\log k})$ where $k=j-i-1$, we can merely replace line~\ref{alg:search_cdf_index} of Alg.~\ref{alg:search_with_distribution} with our function approximator and still enjoy the $O(\log\log n)$ query time. Indeed, the following lemma shows that this is possible using the piecewise approximation of Alg.~\ref{alg:pcf} and under mild assumptions on the data distribution. Let $\hat{r}^{i, j}_t$ be the function returned by \textsc{PCF($A[i+1:j-1]$, $t$, $a_i$, $a_j$)} with $t$ pieces.




\begin{lemma}\label{lemma:prob_func_close}
    Consider two indexes $i, j$, where $1\leq i < j\leq n$ and $a_i<a_j$. Let $k=j-i-1$. For $k\geq 2$, under the conditions of Theorem~\ref{theorem:linear} and for $t\geq\frac{\rho_2}{\rho_1}\sqrt{k}$ we have
$$\mathds{P}(\normx{r^{i, j} -\hat{r}_t^{i, j}}_\infty\geq (\sqrt{0.5\log\log k}+1)\sqrt{k})\leq \frac{1}{\log k}.$$
\end{lemma}

\textit{Proof of Lemma~\ref{lemma:prob_func_close}}. Alg.~\ref{alg:pcf} finds the piecewise constant approximator to $r^{i, j}$ with $t$ pieces with the smallest infinity norm error. Thus, we only need to show the existence of an approximation with $t$ pieces that satisfies conditions of the lemma. To do so, we use the relationship between $r^{i, j}$ and the conditional c.d.f. Intuitively, Lemma~\ref{lemma:dkw} shows that $r^{i, j}$ and the conditional c.d.f are similar to each other and thus, if we can approximate conditional c.d.f well, we can also approximate $r^{i, j}$. Formally, by triangle inequality and for any function approximator $\hat{r}$ we have 
\begin{align}\label{eq:boun:trig}
    \normx{r^{i, j}-\hat{r}}_\infty&\leq \normx{r^{i, j}-kF_\chi^{i, j}}_\infty+\normx{kF_\chi^{i, j}-\hat{r}}_\infty.
\end{align}
Combining this with Lemma~\ref{lemma:dkw} we obtain
$$\mathds{P}(\normx{r^{i, j} -\hat{r}}_\infty\geq \sqrt{0.5k\log\log k}+\normx{kF_\chi^{i, j}-\hat{r}}_\infty)\leq \frac{1}{\log k}.$$

Finally, Lemma~\ref{lemma:appx_cdf} stated below shows how we can approximate the conditional c.d.f and completes the proof. \qed

\begin{lemma}\label{lemma:appx_cdf}
    Under the conditions of Lemma~\ref{lemma:prob_func_close}, there exists a piecewise constant function approximator, $\hat{r}$, with $\frac{\rho_2}{\rho_1}\sqrt{k}$ pieces such that $\normx{\hat{r}-kF_\chi^{i, j}}_\infty\leq \sqrt{k}$.     
%
\end{lemma}


\begin{algorithm}[t]
\begin{algorithmic}[1]
\Require A sorted array $A$ of size $n$ sampled from a distribution $\chi$ with CDF $F_\chi$, a query $q$
\Ensure The root node of the learned index
\Procedure{BuildTree}{$A$, $i$, $j$}
\State $k\leftarrow j-i+1$ \Comment{size of $A[i:j]$}
\If{$k\leq 61$}\label{alg:construc:261}
   \State \Return Leaf node with $\texttt{content}$ $A[i:j]$
\EndIf
\State $\hat{r}, \epsilon\leftarrow$ \textsc{PCF}($A[i:j]$, $\lceil\frac{\rho_2}{\rho_1}\sqrt{k}\rceil$, $a_i$, $a_j$)\label{alg:construct:pcf}
\State $k'\leftarrow \lceil2\sqrt{k}(1+\sqrt{0.5\log\log k})+2\rceil$\label{alg:rda:cover}
\If{$\epsilon>\frac{k'}{2}$ }\label{alg:rda:bound_error}
   \State \Return Leaf node with $\texttt{content}$ $A[i:j]$
\EndIf
\State $C\leftarrow$ array of size $\lceil\frac{k}{k'}\rceil$ containing children
\For{$z\leftarrow 0$ \textbf{to} $\lceil\frac{k}{k'}\rceil$}
    \State $C[z]\leftarrow $ \textsc{BuildTree}($A$, $zk'$, $(z+2)k'$)\label{alg:rda:recursion}
\EndFor
\State \Return Non-leaf node with \texttt{children} $C$ and \texttt{model} $\hat{r}$ with \texttt{max\_err} $\frac{k'}{2}$
\EndProcedure
\caption{RDA Index Construction}\label{alg:construct}
\end{algorithmic}
\end{algorithm}

\begin{figure*}
\begin{minipage}{0.38\textwidth}
    \centering
    \includegraphics[width=\textwidth]{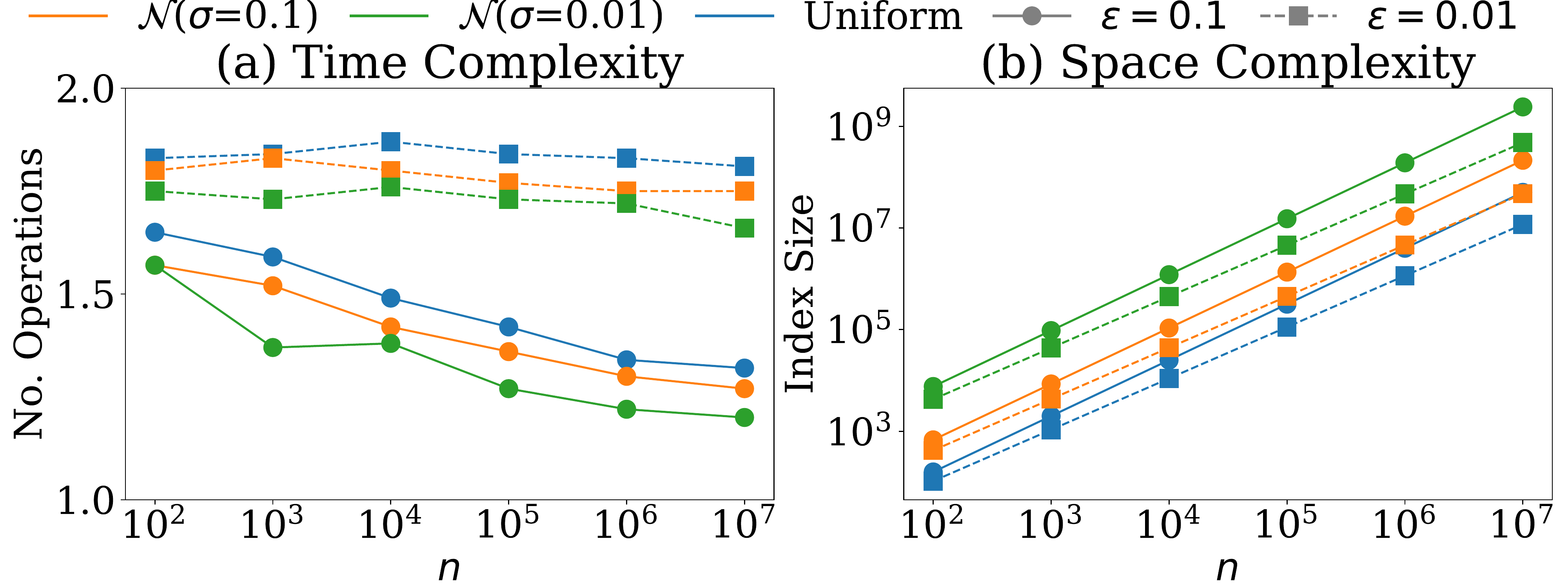}
    \vspace{-0.7cm}
    \caption{Constant Query and Near-Linear Space}
    \label{fig:pca}
\end{minipage}
\hfill
\begin{minipage}{0.2\textwidth}
    \centering
    \includegraphics[width=\textwidth]{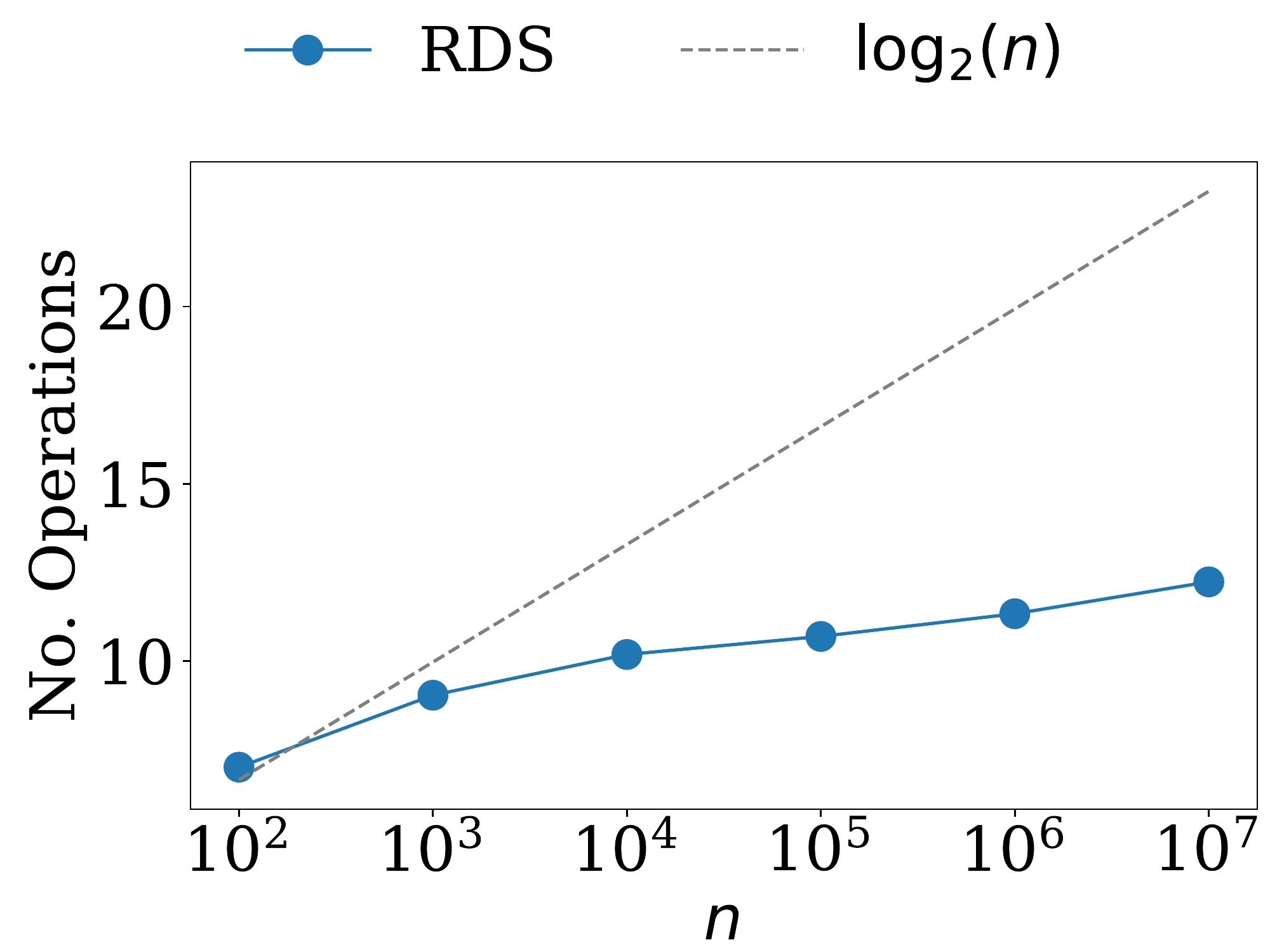}
    \vspace{-0.7cm}
    \caption{Log-Logarithmic Query and Constant Space}
    \label{fig:RDS}
\end{minipage}
\hfill
\begin{minipage}{0.38\textwidth}
    \includegraphics[width=\textwidth]{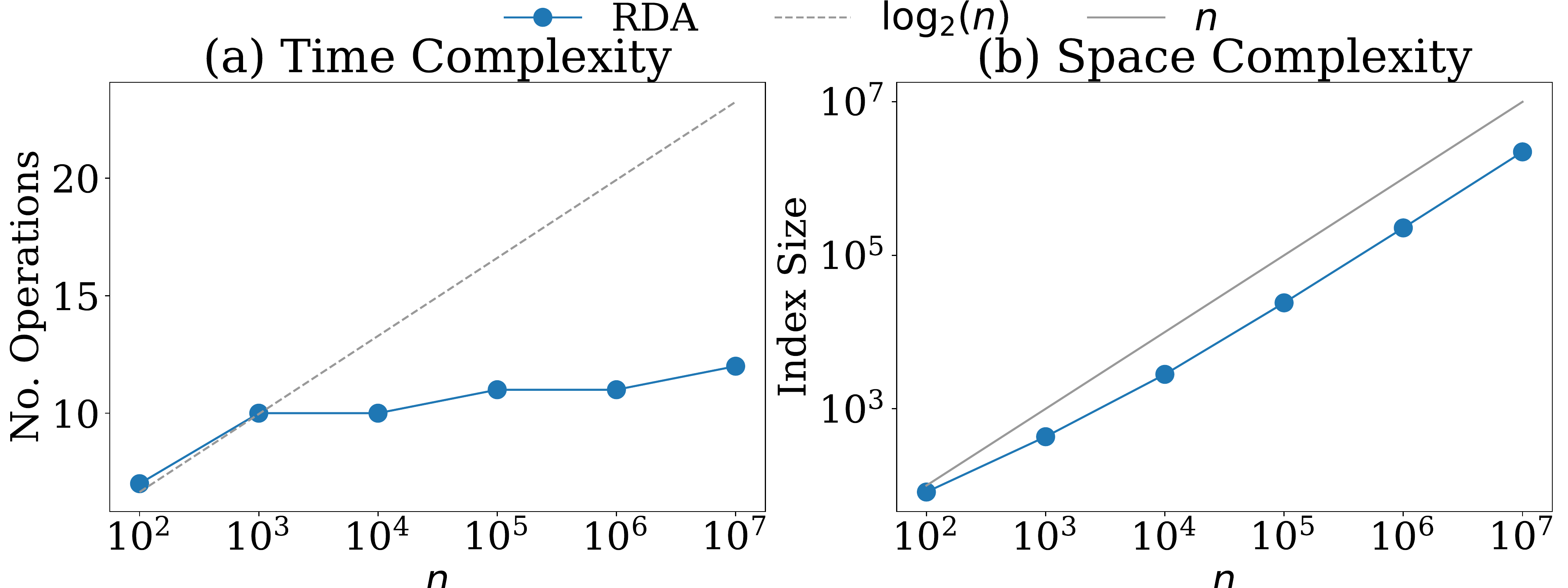}
    \vspace{-0.7cm}
    \caption{Log-Logarithmic Query and Quasi-Linear Space}
    \label{fig:RDA}
\end{minipage}
\end{figure*}

\vspace{-0.2cm}
\subsubsection{Index Construction and Querying}\label{sec:rmi:const}
\vspace{-0.2cm}
Lemma~\ref{lemma:prob_func_close} is an analog of Lemma~\ref{lemma:dkw}, showing a function approximator enjoys similar properties as the c.d.f. However, different function approximators are needed for every subarray (for c.d.f.s we merely needed to scale and shift them differently for different subarrays). Given that there are $O(n^2)$ different subarrays, a naive implementation that creates a function approximator for each subarray takes space quadratic in data size. Instead, we only approximate the conditional rank function for certain sub-arrays while still retaining the $O(\sqrt{k\log\log k})$ error bound per subarray.

\textbf{Construction}. Note that $r(q)=0$ only if $q<a_1$, so we can filter this case out and assume $r(q)\in \{1, ..., n\}$.  RDA is a tree, shown in Fig.~\ref{fig:rmi}, where each node is associated with a model. When querying the index, we traverse the tree from the root, and at each node,  we use the node's model to choose the next node to traverse. Traversing down the tree narrows down the possible answers to $r(q)$. 
We say that a node $N$ \textit{covers} a range $S_N$, if we have $r(q)\in S_N$ for any query, $q$, that traverses the tree and reaches $N$. We call $|S_N|$ node $N$'s \textit{coverage size}. Coverage size is the size of search space left to search after reaching a node. The root node, $N$, covers $\{1, ..., n\}$ with coverage size $n$ and the coverage size decreases as we traverse down the tree. Leaf nodes have coverage size independent of $n$ with high probability, so that finding $r(q)$ takes constant time after reaching a leaf. Each leaf node stores the subarray corresponding to the range it covers as its \texttt{content}.


RDA is built by calling \textsc{BuildTree}($A$, $1$, $n$), as presented in Alg.~\ref{alg:construct}. \textsc{BuildTree}($A$, $i$, $j$) returns the root node, $N$, of a tree, where $N$ covers $\{i, ..., j\}$. 
If the coverage size of $N$ is smaller than some prespecified constant (line~\ref{alg:construc:261}, analogous to line \ref{alg:cdf_search:base1} in Alg.~\ref{alg:search_with_distribution}), the algorithm turns $N$ into a leaf node. Otherwise, in line~\ref{alg:construct:pcf} it uses Lemma~\ref{lemma:prob_func_close} to create the model $\hat{r}$ for $N$, where $\hat{r}$ approximates $r^{i-1, j+1}$ (recall that $r^{i, j}$ is the index function for subarray $A[i+1:j-1]$). If the error of $\hat{r}$ is larger than predicted by Lemma~\ref{lemma:prob_func_close}, the algorithm turns $N$ into a leaf node and discards the model (this is analogous to line \ref{alg:search_cdf_check} in Alg.~\ref{alg:search_with_distribution}). Finally, for $k'$ as in line~\ref{alg:rda:cover}, the algorithm recursively builds $\lceil\frac{k}{k'}\rceil$ children for $N$. Each child has a coverage size of $2k'$ and the ranges are spread at $k'$ intervals (line~\ref{alg:rda:recursion}). This ensures that  the set $\hat{R}=\{\hat{r}-\epsilon, \hat{r}-\epsilon+1, ..., \hat{r}+\epsilon\}$, (with $|\hat{R}|\leq k'$ ensured by line~\ref{alg:rda:bound_error}) is a subset of the range covered by one of $N$'s children. Furthermore, for any query $q$, $\epsilon$ is the maximum error of $\hat{r}$, so $r(q)\in \hat{R}$. Thus, the construction ensures that for any query $q$ that reaches $N$, $r(q)$ is in the range covered by one of the children of $N$. 

 \begin{algorithm}[t]
\begin{algorithmic}[1]
\Require The root node, $N$, of a learned index, a query $q$
\Ensure Rank of query $q$
\Procedure{Query}{$N$, $q$}
\If{$N$ is a leaf node}
    \State \Return \textsc{BinarySearch}($N.$\texttt{content})\label{alg:search_indx:exit}
\EndIf
\State $\hat{i}\leftarrow N.$\texttt{model}$(q)$
\State $k\leftarrow N.$\texttt{max\_err}$(q)$
\State $z\leftarrow \lfloor\frac{\hat{i}-k}{2k}\rfloor$
\State \Return \textsc{Query}($N.\texttt{children}[z]$, $q$)
\EndProcedure
\caption{RDA Index Querying}\label{alg:search_index}
\end{algorithmic}
\end{algorithm}
\textbf{Performing Queries}. As Alg.~\ref{alg:search_index} shows, to traverse the tree for a query $q$ from a node $N$, we find the child of $N$ whose covered range contains $r(q)$. When $\hat{i}$ is $N$.\texttt{model} estimate with maximum error $k$, $z=\lfloor\frac{\hat{i}-k}{2k}\rfloor$ gives the index of the child whose range covers  $\{\lfloor\frac{\hat{i}-k}{2k}\rfloor2k, ..., (\lfloor\frac{\hat{i}-k}{2k}\rfloor+2)2k\}$ and contains $\{\hat{i}-k, ..., \hat{i}+k\}$ as a subset and therefore contains $r(q)$.  Thus, the child at index $z$ is recursively searched.

\subsubsection{Complexity Analysis}\label{sec:linear:proof}
The query time analysis is very similar to the analysis in Sec.~\ref{sec:const_space:analysis} and is thus deferred to Appendix~\ref{appx:proofs}. Here, we show the space overhead complexity. 

All nodes at a given tree level have the same coverage size.  If the coverage size of nodes at level $i$ is $z_i$, then the number of pieces used for approximation per node is $O(\frac{\rho_2}{\rho_1}\sqrt{z_i})$ and the total number of nodes at level $i$ is at most $O(\frac{n}{z_i})$. Thus, total number of pieces used at level $i$ is $c\frac{\rho_2}{\rho_1}\frac{n}{\sqrt{z_i}}$ for some constant $c$.  Note that if the coverage size at level $i$ is $k$, the coverage size at level $i+1$ is $4\sqrt{k}(1+\sqrt{0.5\log\log k})$ which is more than $k^{\frac{1}{2}}$. Thus, $z_i\geq n^{(\frac{1}{2})^i}$ and $c\frac{\rho_2}{\rho_1}\frac{n}{\sqrt{z_i}}\leq c\frac{\rho_2}{\rho_1}\frac{n}{n^{(\frac{1}{2})^{i+1}}}$. The total number of pieces is therefor at most $cn\frac{\rho_2}{\rho_1} \sum_{i=0}^{c'\log\log n}n^{-(\frac{1}{2})^{i+1}}\leq 3cn\frac{\rho_2}{\rho_1}$ for some constant $c'$. Each piece has magnitude $n$ and can be written in $O(\log n)$ bits, so total overhead is $O(\frac{\rho_2}{\rho_1}n\log n)$ bits.\qed

\begin{figure*}
\begin{minipage}{0.79\textwidth}
    \centering
    \includegraphics[width=\textwidth]{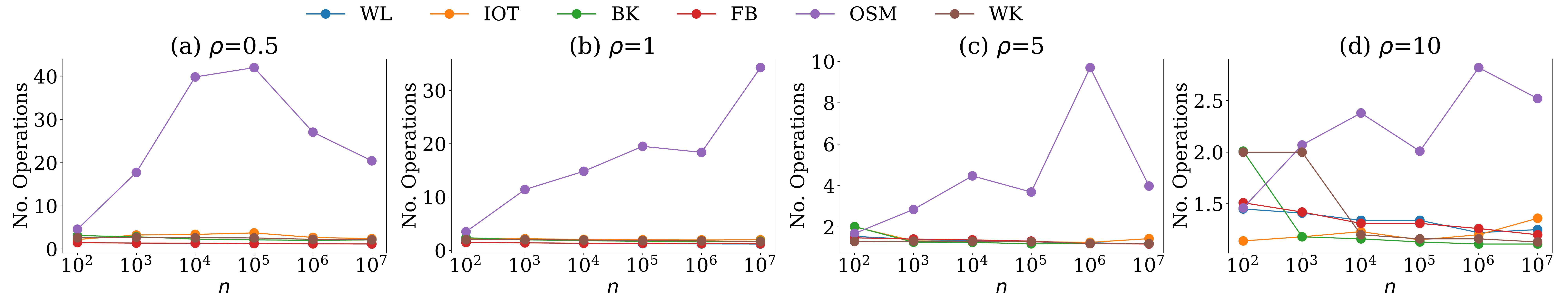}
    \vspace{-0.7cm}
    \caption{Constant Query Time on Real Datasets}
    \label{fig:pca_real_time}
\end{minipage}
\hfill
\begin{minipage}{0.2\textwidth}
    \centering
    \includegraphics[width=\textwidth]{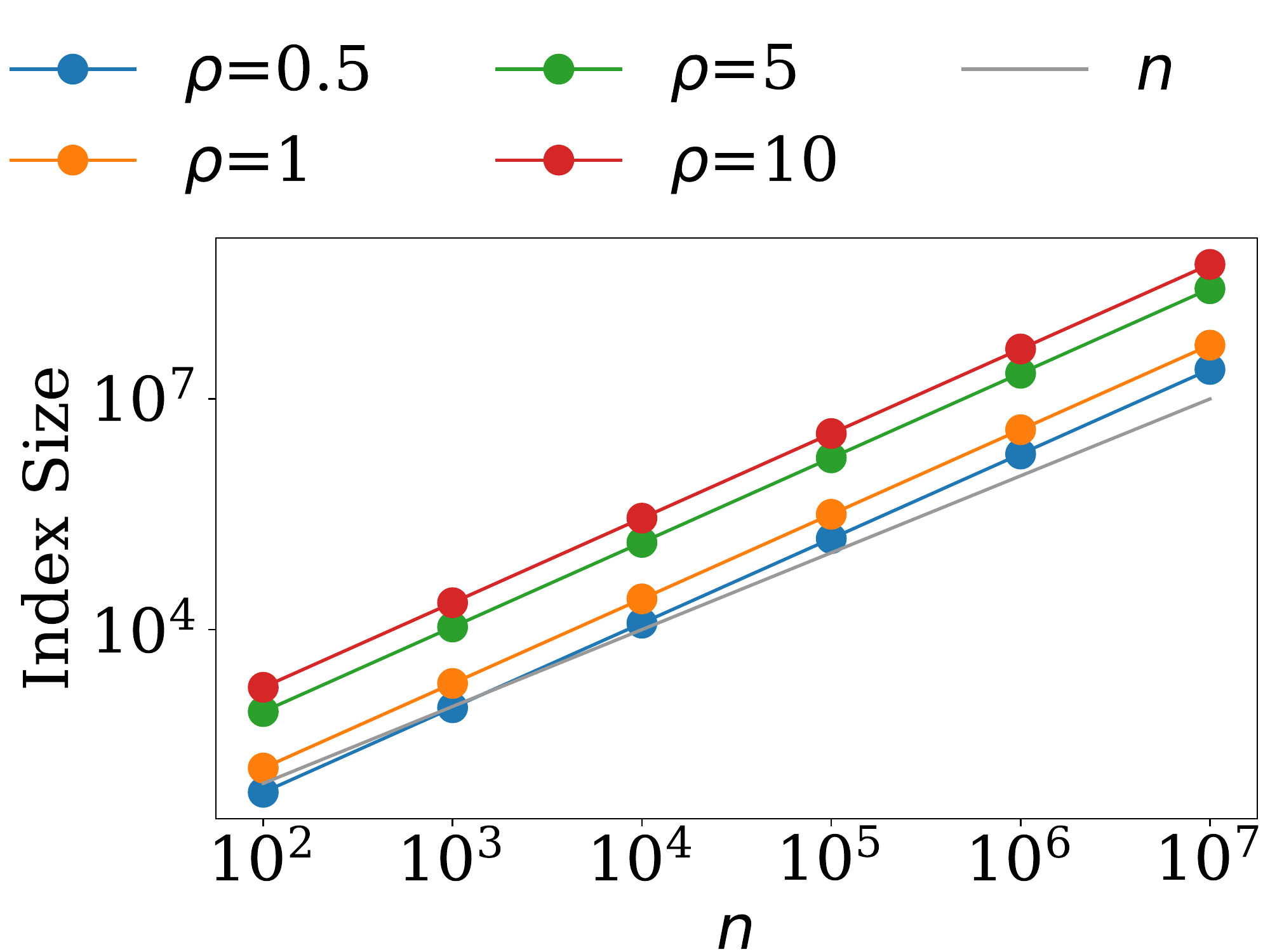}
    \vspace{-0.7cm}
    \caption{Near-Linear Space on Real Datasets}
    \label{fig:pca_real_space}
\end{minipage}
\end{figure*}
\begin{figure*}
\begin{minipage}{0.79\textwidth}
    \centering
    \includegraphics[width=\textwidth]{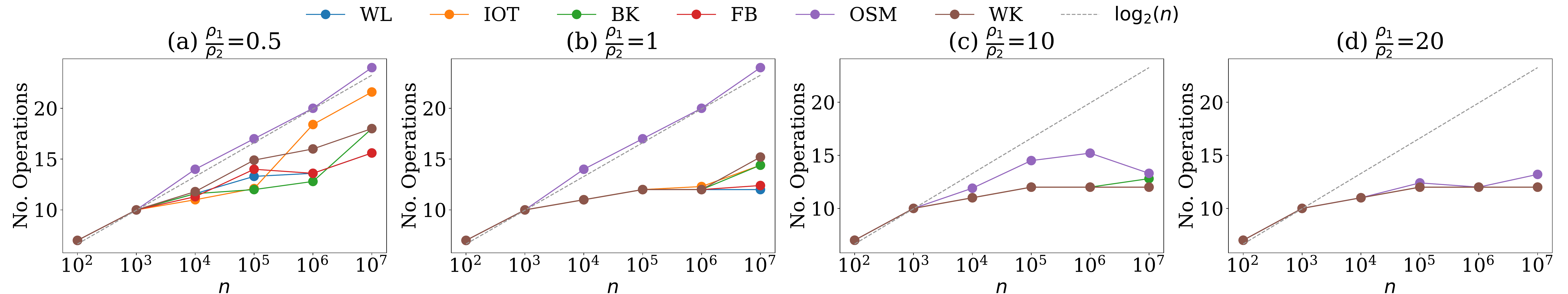}
    \vspace{-0.7cm}
    \caption{Log-Logarithmic Query on Real Datasets}
    \label{fig:rda_real_time}
\end{minipage}
\hfill
\begin{minipage}{0.2\textwidth}
    \centering
    \includegraphics[width=\textwidth]{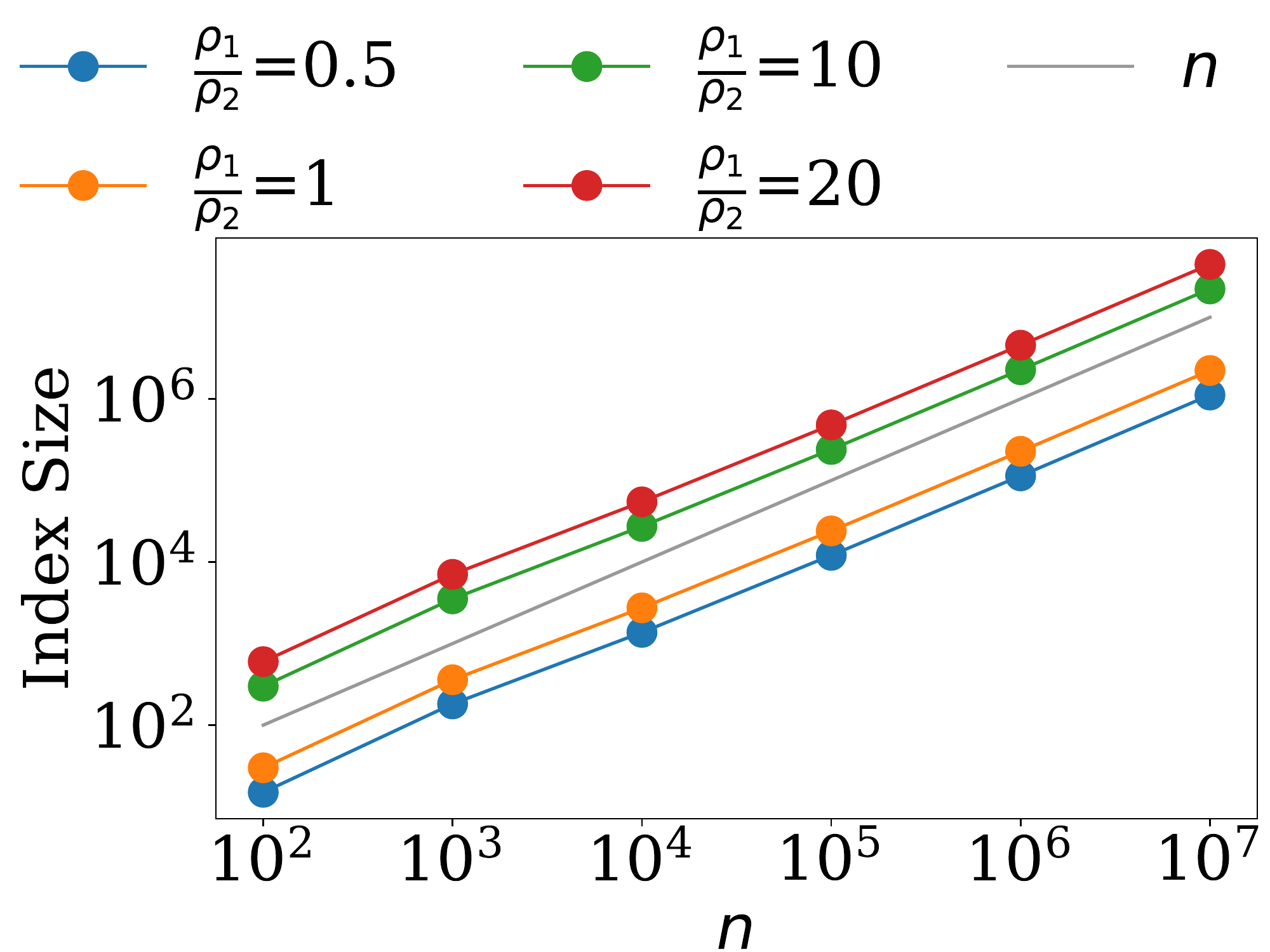}
    \vspace{-0.7cm}
    \caption{Quasi-Linear Space on Real Datasets}
    \label{fig:rda_real_space}
\end{minipage}
\end{figure*}
\vspace{-0.4cm}
\section{Experiments}
\vspace{-0.2cm}
We empirically validate our theoretical results on synthetic 
\revision{and real datasets} (specified in each experiment).

For each experiment, we report \textit{index size} and \textit{number of operations}. Index size is the number of stored integers  by each method. Number of operations is the total number of memory operations performed by the algorithm and is used as a proxy for the total number of instructions performed by CPU. The two metrics differ by a constant factors in our algorithm (our methods perform a constant number of operations between memory accesses), but the latter is compiler dependent and difficult to compute. To report the number of operations, we randomly sample a set of 1000 queries $Q$ and a set of $\mathcal{A}$ of $100$ different arrays from the distribution $\chi$. Let $n_{q, A}$ be the number of operations for each query in $q\in Q$ on an array $A\in \mathcal{A}$. We report $\max_{q\in Q}\frac{\sum_{A\in \mathcal{A}} n_{q, A}}{|\mathcal{A}|}$, which is the maximum (across queries) of the average (across datasets) number of operations. 

\vspace{-0.2cm}
\subsection{Results on Synthetic Datasets}
\vspace{-0.2cm}
\textbf{Constant Query Time and Near-Linear Space}. We show that the construction presented in Sec.~\ref{sec:pcf} achieves the bound of Theorem~\ref{theorem:superlinear}. We consider Uniform and two Gaussian (with $\sigma=0.1$ and $\sigma=0.01$) distributions. We vary Gaussian standard deviation to show the impact of the bound on p.d.f (as required by Theorem~\ref{theorem:superlinear}). Uniform p.d.f. has bound $1$, and bound on Gaussian p.d.f with standard deviation $\sigma$ is $\frac{1}{\sigma\sqrt{2\pi}}$. We present results for $\epsilon=0.1$ and $\epsilon=0.01$, where $\epsilon$ is the space complexity parameter in Theorem~\ref{theorem:superlinear}. 

Fig.~\ref{fig:pca} shows the results. It corroborates Thoerem~\ref{theorem:superlinear}, where Fig.~\ref{fig:pca} (a) shows constant query time achieved by near-linear space shown in Fig.~\ref{fig:pca} (b). We also see for larger $\epsilon$, query time actually decreases, suggesting our bound on query time is less tight for larger $\epsilon$. Furthermore, recall that PCA Index scales the number of pieces by $\rho^{1+\frac{\epsilon}{2}}$ to provide the same bound on query time for all distributions (where $\rho$ is the bound on p.d.f). We see an artifact of this in Fig.~\ref{fig:pca} (b), where when $\rho$ increases index size also increases. 

\textbf{Log-Logarithmic Query Time and Constant Space}. We show that the construction presented in Sec.~\ref{sec:constant_space} achieves the bound of Theorem~\ref{theorem:constant}. The theorem applies to  distributions with efficiently computable c.d.f.s, so we consider distributions over $[0, 1]$ with $F_\chi(x)=x^t$ for $t\in\{1, 4, 16\}$. At $t=1$, we have the uniform distribution and for larger $t$ the distribution becomes more skewed. Fig.~\ref{fig:RDS} corroborates the log-logarithmic bound of Theorem~\ref{theorem:constant}. Moreover, the results look identical across distributions (multiple lines are overlaid on top of each other in the figure), showing similar performance for distributions with different skew levels. 

\textbf{Log-Logarithmic Query Time and Quasi-Linear Space}. We show that the construction presented in Sec.~\ref{sec:linear:proof} achieves the bound of Theorem~\ref{theorem:linear}. We consider Uniform and two Gaussian ($\sigma=0.1$ and $\sigma=0.01$) distributions. The results are presented in Fig.~\ref{fig:RDA}. It corroborates Theorem~\ref{theorem:linear}, where Fig.~\ref{fig:RDA} (a) shows constant query time achieved by quasi-linear space shown in Fig.~\ref{fig:RDA} (b). Similar to the previous case results look identical across distributions (multiple lines are overlaid on top of each other in the figure). Comparing Fig.~\ref{fig:RDA} (a) and Fig.~\ref{fig:RDS}, we observe that using a piecewise function approximator achieves similar results as using c.d.f for rank function approximation.

\vspace{-0.4cm}
\subsection{Results on Real Datasets}\label{sec:exp:real}
\vspace{-0.2cm}
\revision{\textbf{Setup}. On real datasets, we do not have access to the data distribution and thus we do not know the value of $\rho$ in Theorem~\ref{theorem:superlinear} or $\rho_1$ and $\rho_2$ in Theorem~\ref{theorem:linear}. Thus, for each dataset, we perform the experiments for multiple values of $\rho$ and $\rho_1$ or $\rho_2$ to see at what values the trends predicted by the theorems emerge. Since we do not have access to the c.d.f, Theorem~\ref{theorem:constant} is not applicable.}

\revision{We use 6 real datasets commonly used for benchmarking learned indexes. For each real dataset, we sample $n$ data points uniformly at random for different values of $n$ from the original dataset, and queries are generated uniformly at random from the data range. The datasets are WL and IOT from \citet{ferragina2020pgm, kraska2018case, galakatos2019fiting} and BK, FB, OSM, WK from \citet{marcus2020benchmarking} described next. WL: Web Logs dataset containing 714M timestamps of requests to a web server. IOT: timestamps of 26M events recorded by IoT sensors installed throughout an academic building. BK: popularity of 200M books from Amazon, where each key represents the popularity of a particular book. FB: 200M randomly sampled Facebook user IDs, where each key uniquely identifies a user. OSM: cell IDs of 800M  locations from Open Street Map, where each key represents an embedded location. WK: timestamps of 200M edits from Wikipedia, where each key represents the time an edit was committed.}

\revision{\textbf{Results}. Figs.~\ref{fig:pca_real_time} and \ref{fig:pca_real_space} show time and space complexity of the PCA algorithm (Theorem~\ref{theorem:superlinear}) on the real datasets for various values of $\rho$. Note that value of $\rho$ affects the number of pieces used, as described by Lemma~\ref{lemma:const_error}. Furthermore, Figs.~\ref{fig:rda_real_time} and \ref{fig:rda_real_space} show time and space complexity of the RDA algorithm (Theorem~\ref{theorem:linear}) on the real datasets for various ratios of $\frac{\rho_2}{\rho_1}$. Note that value of $\frac{\rho_2}{\rho_1}$ affects the number of pieces used per node, as described by Lemma~\ref{lemma:prob_func_close}. }

\revision{For all except OSM datasets, trends described by Theorems~\ref{theorem:superlinear} and \ref{theorem:linear} hold for values of $\rho$ and $\frac{\rho_2}{\rho_1}$ as small as 1. This shows our theoretical results hold on real datasets, and the distribution dependent factors, $\rho$ and $\frac{\rho_2}{\rho_1}$, are typically small in practice. However, on OSM dataset value of $\rho$ and $\frac{\rho_2}{\rho_1}$ may be as large as 10 and 20 respectively. In fact, \citet{marcus2020benchmarking} shows that non-learned methods outperform learned methods on this dataset. As such, our results provide a possible explanation (large values of $\rho$) for learned methods not performing as well on this dataset. Indeed, OSM is a one-dimensional projection of a spatial dataset using Hilbert curves (see \citet{marcus2020benchmarking}), which distorts the spatial structure of the data and can thus lead to sharp changes to the c.d.f (and therefore large $\rho$). }

\vspace{-0.4cm}
\section{Conclusion}\label{sec:conclusion}
\vspace{-0.2cm}
We theoretically showed and empirically verified that a learned index can achieve sub-logarithmic expected query time under various storage overheads and mild assumptions on data distribution. All our proofs are constructive, using piecewise and hierarchical models that are common in practice. Our results provide evidence why learned indexes perform better than traditional indexes in practice. Future work includes relaxing assumptions on data distribution, finding necessary conditions for sub-logarithmic query time and analyzing the trade-offs of different modeling approaches.  

\vspace{-0.4cm}
\section*{Acknowledgements}
\vspace{-0.2cm}
This research has been funded in part by NSF grants CNS-2125530 and IIS-2128661, and
NIH grant 5R01LM014026. Opinions, findings, conclusions, or recommendations expressed in this material are those of the author(s) and do not necessarily reflect the views of any sponsors, such as NSF.

\bibliography{references}
\bibliographystyle{icml2023}

\appendix
\section{Proofs}\label{appx:proofs}
\textit{Proof of Lemma~\ref{lemma:subexp}}. Assume w.l.o.g that the sub-exponential distribution is centered.  Let $Z_l$ be the event that any point in the array is larger than $l$ or smaller than $-l$ for a positive number $l$. Since the distribution is sub-exponential and using union bound, $\mathds{P}(Z_l)\leq 2ne^{-lK}$ for some constant $K$. To have $2ne^{-lK}\leq\frac{1}{\log n}$ we get that $2n\log n\leq e^{lK}$ and $l\geq\frac{1}{K}\log (2n\log n)$. So let $l=\lceil\frac{2}{k}\log 2n\rceil$ and we have $\mathds{P}(Z_l)\leq \frac{1}{\log n}$. Now, to construct $\hat{R}'$ we first check if any point in $A$ is larger than $l$ or smaller than $-l$. If so, we don't build the index and only do binary search. Otherwise, create $2l$ instances of $\hat{R}$ index, with the $i$-th index covering the range $[-l+i, -l+i+1]$. \revision{Note that interval $[-l+i, -l+i+1]$ has length 1, so that scaling and translating the distribution to interval $[0, 1]$ does not impact the p.d.f of the distribution}. Queries use one of the learned models to find the answer. Thus, the query time is $O(\log n)$ with probability $\frac{1}{\log n}$, and it is $t(n)$ with probability $1-\frac{1}{\log n}$, which is $O(t(n))$. The space overhead is now $O(s(n)\log n)$. \qed

\textit{Proof of Lemma~\ref{lemma:dkw}}. 
\revision{Recall that $X_1$,..., $X_n$ are i.i.d random variables sampled from $\chi$. Furthermore, the array $A$ is a reordering of the random variables so that $A[i] = X_{s_i}$ for some index $s_i$. That is, each element $A[i]$ is itself a random variable and equal to one of $X_1$, ..., $X_n$. $A[i]=a_i$ is a random event.} 

\revision{For $k=j-i-1\geq 2$, let $\{X_{r_{1}}, ..., X_{r_{k}}\}\subseteq \{X_1, ..., X_n\}$ be the elements of the sub-array $A[i+1:j-1]$, which we denote by $X_{r_z}\in A[i+1:j-1]$ for $1\leq z\leq k$. Note that $X_{r_{1}}, ..., X_{r_{k}}$ is not sorted, but the subarray $A[i+1:j-1]$ is the random variables $X_{r_{1}}, ..., X_{r_{k}}$ in a sorted order. For any random variable $X_{r_z}$ for some $z\in \{1, ..., k\}$, we first obtain it's conditional c.f.d given the observations $A[i]=a_i$ and $A[j]=a_j$. The conditional c.d.f can be written as
\begin{align}\label{eq:x_sz_pdf1}
    \mathds{P}(X_{r_z}<x|A[i]=a_i, A[j]=a_j).
\end{align} }

\revision{Let $\bar{X}=\{X_1, ... X_n\}\setminus \{X_{r_z}\}$. Given that $X_{r_z}\in A[i+1:j-1]$, the event $A[i]=a_i, A[j]=a_j, $ is equivalent to the conjunction of the following events: (i) at most $i-1$ of r.v.s in $\bar{X}$ are less than $a_i$, (ii) at least $i$ of r.v.s in $\bar{X}$ are less than or equal to $a_i$, (iii) at most $j-2$ of r.v.s in $\bar{X}$ are less than $a_j$, (iv) at least $j-1$ of r.v.s in $\bar{X}$ are less than or equal to $a_j$, and (v) $a_i\leq X_{r_z}\leq a_j$. This is because (i) and (ii) imply $A[i]=a_i$, while (iii)-(v) imply $A[j]=a_j$. Conversely, $A[i]=a_i$ and $X_{r_z}\in A[i+1:j-1]$ imply (i) and (ii),  $A[j]=a_j$ and $X_{r_z}\in A[i+1:j-1]$ imply (iii) and (iv), and $A[i]=a_i, A[j]=a_j, X_{r_z}\in A[i+1:j-1]$ imply (v). }

\revision{Now, denote by $\phi(\bar{X})$ the event described by (i)-(iv), so that Eq.~\ref{eq:x_sz_pdf1} is 
$$\mathds{P}(X_{r_z}<x|\phi(\bar{X}), a_i\leq X_{r_z}\leq a_j).$$ 
$X_{r_z}$ is independent from all r.v.s in $\bar{X}$, so that Eq.~\ref{eq:x_sz_pdf1} simplifies to }
\revision{$$\mathds{P}(X_{r_z}<x|a_i\leq X_{r_z}\leq a_j)=F_{\chi}^{i, j}(x),$$}
\revision{For all $X_{r_z}\in \{X_{r_{1}}, ..., X_{r_{k}}\}$. Thus, the r.v.s in $\{X_{r_{1}}, ..., X_{r_{k}}\}$ have the conditional c.d.f $F_{\chi}^{i, j}(x)$.} 

\revision{A similar argument to the above shows that r.v.s in any subset of $X_{r_{1}}, ..., X_{r_{k}}$ are independent, given $A[i]=a_i, A[j]=a_j$. Specifically, let $\Tilde{X}\subseteq \{X_{r_{1}}, ..., X_{r_{k}}\}$ for $|\Tilde{X}|\geq 2$. Then, the joint c.d.f of r.v.s in $\Tilde{X}$ can be written as}

\revision{\begin{align}\label{eq:x_sz_pdf2}
    P(\forall_{X\in \Tilde{X}} X<x|&A[i]=a_i, A[j]=a_j).
\end{align}}

\revision{Similar to before, define $\bar{X}=\{X_1, ..., X_n\}\setminus\Tilde{X}$. Given that $\forall_{X\in \Tilde{X}}X\in A[i+1:i-1]$, the event $A[i]=a_i, A[j]=a_j$  is equivalent to the conjunction of the following events: (i) at most $i-1$ of r.v.s in $\bar{X}$ are less than $a_i$, (ii) at least $i$ of r.v.s in $\bar{X}$ are less than or equal to $a_i$, (iii) at most $j-1-|\Tilde{X}|$ of r.v.s in $\bar{X}$ are less than $a_j$, (iv) at least $j-|\Tilde{X}|$ of r.v.s in $\bar{X}$ are less than or equal to $a_j$, and (v) $\forall X\in\Tilde{X}$, we have $a_i\leq X\leq a_j$. Now let $\phi(\bar{X})$ be the event described by (i)-(iv), so that Eq.~\ref{eq:x_sz_pdf2} can be written as 
$$P(\forall_{X\in \Tilde{X}}X<x|\phi(\bar{X}), \forall_{X\in \Tilde{X}} a_i\leq X\leq a_j).$$ 
All r.v.s in $\Tilde{X}$ are independent from all r.v.s in $\bar{X}$, so that Eq.~\ref{eq:x_sz_pdf2} simplifies to 
$$P(\forall_{X\in \Tilde{X}}X<x|\forall_{X\in \Tilde{X}} a_i\leq X\leq a_j).$$
Finally, all r.v.s in $\Tilde{X}$ are also independent from each other, so we obtain that Eq.~\ref{eq:x_sz_pdf2} is equal to 
\begin{align}\label{eq:x_sz_pdf3}
    \Pi_{X\in \Tilde{X}} P(X<x|a_i\leq X\leq a_j),
\end{align}
Proving the independence of r.v.s in $\Tilde{X}$ conditioned on $A[i]=a_i, A[j]=a_j$.}




\revision{To summarize, we have shown that $X_{r_1}, ..., X_{r_k}$ r.v.s conditioned on $A[i]=a_i, A[j]=a_j$ are $k$ i.i.d random variables with the c.d.f $F_{\chi}^{i, j}(x)$}. Moreover, $\frac{1}{k}r_A^{i, j}(x)$ is the empirical c.d.f of the $k$ r.v.s. By DKW bound \cite{massart1990tight} and for $t\geq0$, we have
\begin{align*}
\mathds{P}(\sup_{x}|\frac{1}{k}r_A^{i, j}(x)-F^{i, j}_\chi(x)|\geq \frac{t}{\sqrt{k}})\leq 2e^{-2t^2}.
\end{align*}
Rearranging and substituting $t=\sqrt{0.5\log\log k}$ proves the lemma.\qed

\textit{Proof of Lemma~\ref{lemma:appx_cdf}}. 
$k=j-i-1$. Divide the range $[a_i, a_j]$ into $t$ uniformly spaced pieces, so that the $z$-th piece approximates $kF_\chi^{i, j}$ over $I_z=[a_i + z\frac{a_j-a_i}{t}, a_i + (z+1)\frac{a_j-a_i}{t}]$, which is an interval of length $\frac{a_j-a_i}{t}$. Let $P(x)$ be the constant in the Taylor expansion of $kF_\chi^{i, j}$ around some point in $I_z$. By Taylor's remainder theorem, 
\begin{align}\label{eq:taylor}
\sup_{x\in I_z}|P(x)-kF_{\chi}^{i, j}(x)|\leq k\times \frac{f_\chi(c)}{F_\chi(a_j)-F_\chi(a_i)}\times\frac{a_j-a_i}{t}    
\end{align}
for some $c\in I_z$, where we have used the fact that the derivative of the c.d.f is the p.d.f, and that any two point in $I_z$ are at most $\frac{a_j-a_i}{t}$ apart. 

By mean value theorem, there exist a $c'\in I_z$ so that $\frac{F_\chi(a_j)-F_\chi(a_i)}{a_j-a_i}=f_\chi(c')$. This together with Eq.~\ref{eq:taylor} yields
$$\sup_{x\in R}|P(x)-kF_{\chi}^{i, j}(x)|\leq k\times\frac{f_\chi(c)}{f_\chi(c')}\times\frac{1}{t}\leq \frac{\rho_2}{\rho_1}\frac{k}{t}.$$
Setting $t\geq\frac{\rho_2}{\rho_1}\sqrt{k}$ ensures $\frac{\rho_2}{\rho_k}\frac{k}{t}\leq \sqrt{k\log\log k}$ so that 

\begin{align*}
    \normx{P(x)-kF_\chi^{i, j}(x)}_\infty\leq \sqrt{k\log\log k}
\end{align*}
\qed

\textit{Proof of Query Time Complexity of Theorem~\ref{theorem:linear}}. The algorithm traverses the tree recursively proceeds. At each recursion level, the algorithm performs a constant number of operations unless it perform a binary search. Let the depth of recursion be $h$ and let $k_i$ be the coverage size of the node at the $i$-th level of recursion (so that binary search at $i$-th level takes $O(\log k_i)$). Let $B_i$ denote the event that the algorithm performs binary search at the $i$-th iteration. Thus, for any query $q$, the expected number of operations is $$\mathds{E}_{A\sim \chi}[T(\hat{r}, q)]=\sum_{i=1}^{h} c_1+c_2\mathds{P}(B_i,\bar{B}_{i-1}, .... \bar{B}_1)\log k_i$$ for constants $c_1$ and $c_2$. Note that 
$\mathds{P}(B_i,\bar{B}_{i-1}, .... \bar{B}_1)\leq \mathds{P}(B_i|\bar{B}_{i-1}, .... \bar{B}_1)$, 
where $\mathds{P}(B_i|\bar{B}_{i-1}, .... \bar{B}_1)$ is the probability that the algorithm reaches $i$-th level of tree and performs binary search. By Lemma~\ref{lemma:prob_func_close}, this probability bounded by $\frac{1}{\log k_i}$. Thus $\mathds{E}_{A\sim \chi}[T(\hat{r}, q)]$ is $O(h)$.

To analyze the depth of recursion, recall that at the last level, the size of the array is at most 61. Furthermore, at every iteration the size of the array is reduced by $4(\sqrt{n})(1+\sqrt{0.5  \log\log n})$. For $n\geq 61$, $4(\sqrt{n})(1+\sqrt{0.5  \log\log n})\leq n^c$ for some constant $c<1$, so that the size of the array at the $i$-th recursions is at most $n^{c^i}$ and the depth of recursion is $O(\log\log n)$. Thus, the expected total time is $O(\log\log n)$\qed.

%

\textit{Proof of Prop.~\ref{prop:bound_err_by_samples}}. Note that $r_A(x)$ is a non-decreasing step function, where each step has size 1. Let $s_i$ be the number of steps of $r_A(x)$ in the interval $I_i$. Therefore, 
\begin{align}\label{eq:bound_by_no_steps}
    |r_A(x)-r_A(x')|\leq s_i,
\end{align} for any $x,x'\in I_i$.  Therefore, for $x\in I_i$, substituting $\hat{r}(x;\theta)=r_A(i\times \frac{1}{k})$ into Eq.~\ref{eq:bound_by_no_steps} we get  
\begin{align}\label{eq:bound_by_no_steps2}
    e_i\leq s_i.
\end{align}
Furthermore, points of discontinuity (i.e., steps) of $r_A(x)$ occur when $x=A[j]$ for $j\in [n]$. Therefore, $s_i=|\{j|A[j]\in I_i\}|$. That is, $s_i$ is equal to the number of points in $A$ that are sampled in the range $I_i$. \qed

\textit{Proof of Lemma~\ref{lemma:prob_poins_in_piece}}. Specifically, we bound the probability that $s_{max}\geq c$, for some constant $c$. In other words, we bound the probability of the event, $E$, that any interval has more than $c$ points sampled in it, for any $c\geq 3$. Let $\delta_i=F_{\chi}(\frac{i+1}{k})-F_{\chi}(\frac{i}{k})$ be the probability that a point falls inside $I_i$, so that $\delta_i^c$ is the probability that a set of $c$ sampled points fall inside $I_i$. Taking the union bound over all possible subsets of size $c$, we get that the probability, $p_i$, that the $i$-th interval has $c$ points or more is at most
$$
p_i\leq {n \choose c} \delta_i^c\leq \frac{(en)^c}{c^c}\delta_i^c,
$$
Where the second inequality follows from Sterling's approximation. By mean value theorem, there exists $c\in [\frac{i}{k}, \frac{i+1}{k}]$ such that $F_\chi(\frac{i+1}{k})-F_\chi(\frac{i}{k})=f_\chi(c)(\frac{1}{k})$. Therefore, $\delta_i\leq \frac{\rho}{k}$. Thus, by union bound

\begin{align}\label{eq:err_p}
P(E) \leq \sum_{i=1}^{k}\frac{(en)^c}{c^c}(\frac{\rho}{k})^c=k\frac{(en)^c}{c^c}(\frac{\rho}{k})^c.
\end{align}
Now set $k\geq n^{1+\frac{2}{c-1}}\rho^{1+\frac{1}{c-1}}$ and substitute into Eq.~\ref{eq:err_p}, we obtain that $P(E)\leq \frac{1}{n}$.  \qed


\end{document}